\documentclass[aps,prl,tightenlines,superscriptaddress,showpacs,byrevtex]{revtex4}
\usepackage{graphicx} 
\usepackage{dcolumn}  
\graphicspath{{ps}}

\begin{document}


\title{\quad\\[0.5cm]  Study of 
$ B^- \to D^{**0}\pi^-(D^{**0}\to D^{(*)+}\pi^-)$ decays}

\tighten
\affiliation{Budker Institute of Nuclear Physics, Novosibirsk}
\affiliation{University of Cincinnati, Cincinnati, Ohio 45221}
\affiliation{University of Frankfurt, Frankfurt}
\affiliation{Gyeongsang National University, Chinju}
\affiliation{University of Hawaii, Honolulu, Hawaii 96822}
\affiliation{High Energy Accelerator Research Organization (KEK), Tsukuba}
\affiliation{Hiroshima Institute of Technology, Hiroshima}
\affiliation{Institute of High Energy Physics, Chinese Academy of Sciences, Beijing}
\affiliation{Institute of High Energy Physics, Vienna}
\affiliation{Institute for Theoretical and Experimental Physics, Moscow}
\affiliation{J. Stefan Institute, Ljubljana}
\affiliation{Kanagawa University, Yokohama}
\affiliation{Korea University, Seoul}
\affiliation{Kyungpook National University, Taegu}
\affiliation{Institut de Physique des Hautes \'Energies, Universit\'e de Lausanne, Lausanne}
\affiliation{University of Ljubljana, Ljubljana}
\affiliation{University of Maribor, Maribor}
\affiliation{University of Melbourne, Victoria}
\affiliation{Nagoya University, Nagoya}
\affiliation{Nara Women's University, Nara}
\affiliation{National Kaohsiung Normal University, Kaohsiung}
\affiliation{National Lien-Ho Institute of Technology, Miao Li}
\affiliation{Department of Physics, National Taiwan University, Taipei}
\affiliation{H. Niewodniczanski Institute of Nuclear Physics, Krakow}
\affiliation{Nihon Dental College, Niigata}
\affiliation{Niigata University, Niigata}
\affiliation{Osaka City University, Osaka}
\affiliation{Osaka University, Osaka}
\affiliation{Panjab University, Chandigarh}
\affiliation{Peking University, Beijing}
\affiliation{RIKEN BNL Research Center, Upton, New York 11973}
\affiliation{Saga University, Saga}
\affiliation{University of Science and Technology of China, Hefei}
\affiliation{Seoul National University, Seoul}
\affiliation{Sungkyunkwan University, Suwon}
\affiliation{University of Sydney, Sydney NSW}
\affiliation{Tata Institute of Fundamental Research, Bombay}
\affiliation{Toho University, Funabashi}
\affiliation{Tohoku Gakuin University, Tagajo}
\affiliation{Tohoku University, Sendai}
\affiliation{Department of Physics, University of Tokyo, Tokyo}
\affiliation{Tokyo Institute of Technology, Tokyo}
\affiliation{Tokyo Metropolitan University, Tokyo}
\affiliation{Tokyo University of Agriculture and Technology, Tokyo}
\affiliation{Toyama National College of Maritime Technology, Toyama}
\affiliation{University of Tsukuba, Tsukuba}
\affiliation{Utkal University, Bhubaneswer}
\affiliation{Virginia Polytechnic Institute and State University, Blacksburg, Virginia 24061}
\affiliation{Yokkaichi University, Yokkaichi}
\affiliation{Yonsei University, Seoul}
  \author{K.~Abe}\affiliation{High Energy Accelerator Research Organization (KEK), Tsukuba} 
  \author{K.~Abe}\affiliation{Tohoku Gakuin University, Tagajo} 
  \author{T.~Abe}\affiliation{Tohoku University, Sendai} 
  \author{I.~Adachi}\affiliation{High Energy Accelerator Research Organization (KEK), Tsukuba} 
  \author{H.~Aihara}\affiliation{Department of Physics, University of Tokyo, Tokyo} 
  \author{M.~Akatsu}\affiliation{Nagoya University, Nagoya} 
  \author{Y.~Asano}\affiliation{University of Tsukuba, Tsukuba} 
  \author{T.~Aso}\affiliation{Toyama National College of Maritime Technology, Toyama} 
  \author{T.~Aushev}\affiliation{Institute for Theoretical and Experimental Physics, Moscow} 
  \author{A.~M.~Bakich}\affiliation{University of Sydney, Sydney NSW} 
% ...
\collaboration{The Belle Collaboration}

\collaboration{The Belle Collaboration}

\noaffiliation

\begin{abstract}

 We report the results of a study of charged $B$ decays to the
$D^{\pm}\pi^{\mp}\pi^{\mp}$ and $D^{*\pm}\pi^{\mp}\pi^{\mp}$ final states 
using complete $D^{(*)}$ reconstruction. The contributions of two-body
$B \to D^{**}\pi$ decays with narrow (j=3/2) and broad (j=1/2)
$D^{**}$  states have been determined and the 
masses and widths of four $D^{**}$ states
have been measured. This is the first observation of the broad
$D^{*0}_0$ and $D'^{0}_1$ mesons.
The analysis is based on a  data sample of 65 million $B\bar{B}$ pairs
collected in the Belle experiment. 

\end{abstract}


\maketitle
       
\tighten

{\renewcommand{\thefootnote}{\fnsymbol{footnote}}}
\setcounter{footnote}{0}

\section{Introduction}

\begin{figure}[h]
\begin{center}
\includegraphics[height=12 cm, width=12 cm]{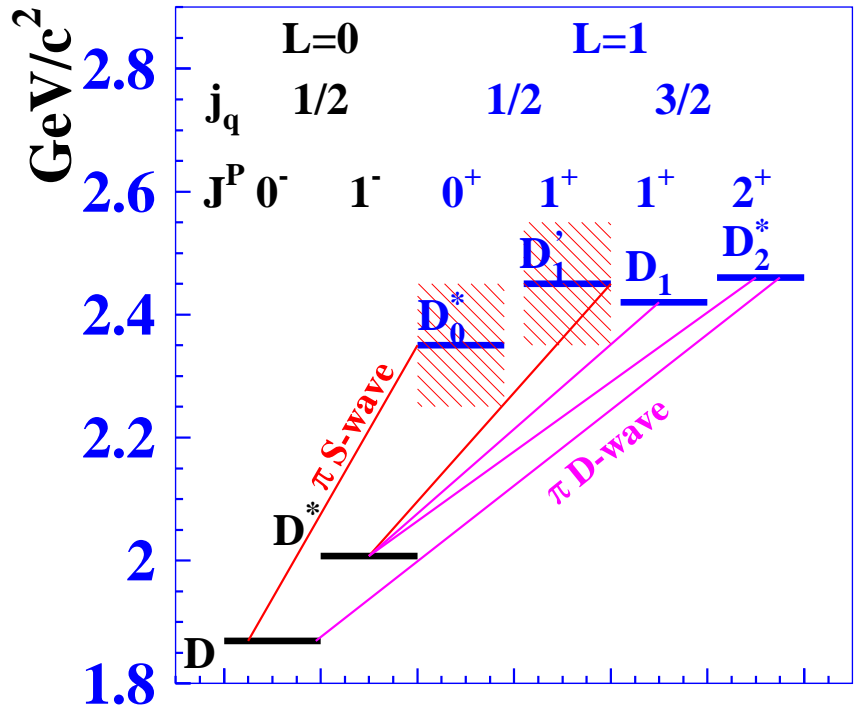}
\end{center}

\vspace*{-2. cm}
\caption{Spectroscopy of $D$-meson excitations. The lines show
  possible single
 pion 
transitions.}
\label{fi:spec}
\end{figure}

$B$ decays
to $D\pi$ and $D^*\pi$ final states are two of the 
dominant  hadronic $B$ decay modes and have been measured quite 
well~\cite{PDG}.
In this paper we study the   production of $D$-meson excited states,
collectively refered to as
$D^{**}$'s, that are P-wave excitations of quark-antiquark  systems 
containing one charmed and one light ($u,d$) quark. 
The results provide tests of Heavy Quark Effective Theory (HQET) and
QCD sum rules.
Figure~\ref{fi:spec} shows the spectroscopy of  
$D$-meson excitations. In the heavy quark limit, the heavy quark spin
${\vec s}_c$ decouples from the other degrees of freedom and the total
angular momentum of the light quark ${\vec j}_q=\vec{L}+{\vec s}_q$ is a good 
quantum number.
There are four P-wave states with the following spin-parity and light 
quark angular  momenta:
$0^+(j_q=1/2),~1^+(j_q=1/2),~1^+(j_q=3/2)$ and $2^+(j_q=3/2)$, which are 
usually labeled as $D^*_0,~D'_1,~D_1$ and $D^*_2$, respectively.   
The two $j_q=3/2$ states are narrow with widths of order 20~MeV
and have already been 
observed~\cite{AR1,AR2,AR3,e691,CL15,e687,CL2,dobs,DELPHI,DELPHI1,ALEPH}.
 The measured values of their masses 
agree with model predictions~\cite{isgur,rosner,godfrey,falk}.
The remaining $j_q=1/2$ states decay via S-waves and are expected to be 
quite broad. Although they have not yet been directly observed,  their 
total production rate has been measured in $B$-meson semileptonic 
decays~\cite{DELPHI}.
 
CLEO has observed the production of both of the narrow $D^{**}$ mesons in  
$B\to D^{*}\pi\pi$ decays with the following branching 
fractions~\cite{CLEO9625}:
\begin{eqnarray}
{\cal B}(B^-\to D^0_1\pi^-)\times {\cal B}(D_1^0\to D^{*+}\pi^-)&=&(7.8\pm1.9)\times10^{-4},\nonumber\\
{\cal B}(B^-\to D^{*0}_2\pi^-)\times {\cal B}(D_2^{*0}\to D^{*+}\pi^-)&=&(4.2\pm1.7)\times10^{-4}.\label{e:cl}
\end{eqnarray}
The ratio of the $B$ meson branching fractions
\begin{equation}
\label{e:Neu1}
R=\frac{{\cal B}(B^-\to D^{*0}_2\pi^-)}{{\cal B}(B^-\to D^0_1\pi^-)}
\end{equation}
 is calculated 
in HQET and the factorization approach  in Refs.~\cite{leg,neubert}.
In Ref.~\cite{leg} 
$R$ is found to depend on
the values of  
sub-leading Isgur-Wise functions ($\hat{\tau}_{1,2}$)
describing $\Lambda_{QCD}/m_c$ corrections. 
Variations of $\hat{\tau}_{1,2}$ by  $\pm0.75$~GeV result in
values of $R$ that range from 0 to 1.5.
In  Ref.~\cite{neubert} some of the sub-leading terms are estimated
and the ratio is determined to be  
\begin{equation}
\label{e:Neu}
R\approx 0.35\biggl|\frac{1+\delta_8^{D2}}{1+\delta_8^{D1}}\biggr|^2,
\end{equation}
where $\delta_8^{D1(D2)}$ are non-factorized corrections that are expected 
to be small. 
The value of $R$ calculated from the CLEO 
results given in Eq.(\ref{e:cl}) plus the ratio of branching fractions 
 ${B(D^{*0}_2\to D^{+}\pi^-)}/{B(D^{*0}_2\to
  D^{*+}\pi^-)}=2.3\pm0.8$~\cite{CL2,AR3} and the assumption that
$D_1$ and $D_2^*$ decays are saturated by the two-body $D\pi,~D^*\pi$
modes,
 is
$R=1.8\pm0.8$. This value  is higher than the prediction,  
although the  uncertainties are large.
 If more precise measurements 
do not indicate lower values of $R$, a problem for theory may arise.
Thus, a measurement of $R$ will allow us to test
HQET predictions.

Another possible inconsistency between theory and experiment 
is in the ratio of the production rates of 
narrow and broad states in semileptonic $B$ decays.
QCD sum rules~\cite{QCDSR}
predict the dominance of
narrow $D^{**}$($j_q=3/2$) state production in $B\to D^{**}l\nu$ decays.
On the other hand,  
the total branching fraction 
${\cal B}(B\to D^{(*)}\pi l^-\bar{\nu})=(2.6\pm0.5)\%$
measured by ALEPH and DELPHI~\cite{PDG}
is not saturated by the contribution of the narrow resonances, 
$(0.86\pm0.37)\%$~\cite{cleol},   indicating a large 
contribution of   broad or
nonresonant $D^{(*)}\pi$ structures.

In this study we concentrate on charged $B$ decays
to $D^{(*)\pm}\pi^{\mp}\pi^{\mp}$. For these decays the  final state contains 
two pions of the same sign that do not form any bound states,
making analysis of the final state simpler.  

\section{The Belle detector}

  The Belle detector~\cite{Belle} is a large-solid-angle magnetic spectrometer
that consists
of a three-layer silicon vertex detector (SVD), 
a 50-layer central drift chamber
(CDC) for charged particle tracking and specific ionization measurement 
($dE/dx$), an array of aerogel threshold \v{C}erenkov counters (ACC), 
time-of-flight scintillation counters (TOF), and an array of 8736 CsI(Tl) 
crystals for electromagnetic calorimetry (ECL) located inside a superconducting
solenoid coil that provides a 1.5~T magnetic field. An iron flux return located
outside the coil is instrumented to detect $K_L$ mesons and identify muons
(KLM). 
We use a GEANT-based Monte Carlo (MC) simulation to
model the response of the detector and determine the acceptance~\cite{sim}.

  Separation of kaons and pions is accomplished by combining the responses of 
the ACC and the TOF with $dE/dx$ measurements in the CDC
 to form a likelihood $\cal{L}$($h$) where $h=(\pi)$ or $(K)$. 
Charged particles 
are identified as pions or kaons using the likelihood ratio 
(PID):
\[{\rm PID}(K)=\frac{{\cal{L}}(K)}{{\cal{L}}(K)+{\cal{L}}(\pi)};~~
{\rm PID}(\pi)=
\frac{{\cal{L}}(\pi)}{{\cal{L}}(K)+{\cal{L}}(\pi)}=1-{\rm PID}(K).\]
  At large momenta ($>$2.5 GeV/$c$) only the ACC and $dE/dx$ are used since the
TOF provides no significant separation of kaons and pions. 
Electron identification is based on a combination of $dE/dx$ measurements,
the ACC responses  and the position, shape and total energy deposition
($E/p$) of the shower detected in the ECL. A more detailed
description of the Belle particle identification can be found in 
ref.~\cite{PID}.

\section{Event selection}
A 60.4~fb$^{-1}$ data sample 
(65.4 million~$B\bar{B}$ events) collected at the 
$\Upsilon (4S)$ resonance with the Belle detector is used.
Candidate $B^-\to D^+\pi^-\pi^-$ and $B^-\to D^{*+}\pi^-\pi^-$ events
as well as charge conjugate combinations are selected.
The $D^+$ and $D^{*+}$ mesons are reconstructed in the 
$D^+\to K^-\pi^+\pi^+$ and $D^{*+}\to D^0\pi^+$ modes, respectively.
The
$D^0$ candidates are reconstructed in the $D^0\to K^-\pi^+$ and
$D^0\to K^-\pi^+\pi^+\pi^-$ channels. 
The signal-to-noise ratios for other $D$ decay modes are found to be much
lower and they are not used in this analysis.

Charged tracks are selected with requirements based on the  
average hit residuals and impact parameters relative to the interaction
point. We also require that the polar angle of each track be within
the angular  range of $17^{\circ}-150^{\circ}$ and that the transverse track 
momentum be 
greater than 50 MeV/c for kaons and 25 MeV/c for pions.

Charged kaon candidates are selected with the requirement ${\rm PID}(K)>0.6$. 
This  has an efficiency of
$90\%$ for kaons
and a pion misidentification probability of $10\%$.
For pions the requirement \mbox{${\rm PID}(\pi)>0.2$} is used.
All tracks that are positively identified as electrons are rejected. 

$D^{+}$ mesons are reconstructed from  
$K^{-}\pi^{+}\pi^{+}$ combinations with
invariant mass within 13~MeV/c$^2$ of the nominal $D^+$ mass, which
corresponds to about 3\,$\sigma_{K\pi\pi}$. For ${D}^0$ mesons,
the $K\pi$ or $K\pi\pi\pi$ invariant mass  is required to be within 
15~MeV/c$^2$ of the nominal $D^0$ mass 
(3\,$\sigma_{K\pi}$).
We reconstruct $D^{*+}$ mesons from the $D\pi$ combinations 
with a mass difference of $M_{D\pi}-M_{D^0}$ within $1.5~{\rm
  MeV}/c^2$ of its nominal value.

Candidate events are identified 
by their center of mass (c.m.)\ energy difference, 
$\Delta E=(\sum_iE_i)-E_{\rm b}$, and 
beam-constrained mass, 
$M_{\rm bc}=\sqrt{E^2_{\rm b}-(\sum_i\vec{p}_i)^2}$, where 
$E_{\rm b}=\sqrt{s}/2$ is the beam energy in the $\Upsilon(4S)$
c.m.\ frame, and $\vec{p}_i$ and $E_i$ are the c.m.\ three-momenta and 
energies of the $B$ meson candidate decay products. We select events with 
$M_{\rm bc}>5.20$~GeV/$c^2$ and $|\Delta E|<0.10$~GeV.

To suppress the large continuum background ($e^+e^-\to q\bar{q}$,
where $q=u,d,s,c$), topological variables are used. Since  
the produced $B$ mesons 
are almost at rest in the c.m. frame, the angles of the decay products
of the two $B$ mesons are uncorrelated and the tracks tend to be 
isotropic while  continuum $q\bar{q}$ events
tend to have  a two-jet structure. We use the angle between the thrust axis of 
the $B$ candidate and that of the rest of the event ($\Theta_{thrust}$)
to discriminate between these two cases. The distribution of
$|\cos\Theta_{thrust}|$ is strongly peaked near $|\cos\Theta_{thrust}|=1$
for $q\bar{q}$ events and is nearly flat for  $\Upsilon(4S)\to
B\bar{B}$ events.
We require $|\cos\Theta_{thrust}|<0.8$, which eliminates about 
83$\%$ of the continuum background while retaining about 80$\%$ of signal 
events.

There are events for which two or more combinations pass all
the selection criteria. According to a MC simulation,
this occurs  primarily
because of the misreconstruction of a low momentum pion from the 
$D^{**}\to D^{(*)}\pi$ decay. To avoid multiple entries, 
the combination that has the  minimum difference of 
$Z$ coordinates at the interaction point, $|Z_{\pi_1}-Z_{\pi_2}|$, 
of the tracks corresponding to the pions from 
$B\to D^{**}\pi_1$ and $D^{**}\to D^{(*)}\pi_2$ decays
is selected~\cite{foot1}.
 This selection 
suppresses the combinations that
include pions from $K_S$ decays. In the case of multiple $D$ 
combinations, the one with invariant
mass closest to the nominal value is selected.  
 
\section{$B^{-}\to D^{+}\pi^{-}\pi^{-}$ analysis}

\begin{figure}[h]
\begin{tabular}{cc}

\includegraphics[height=6 cm]{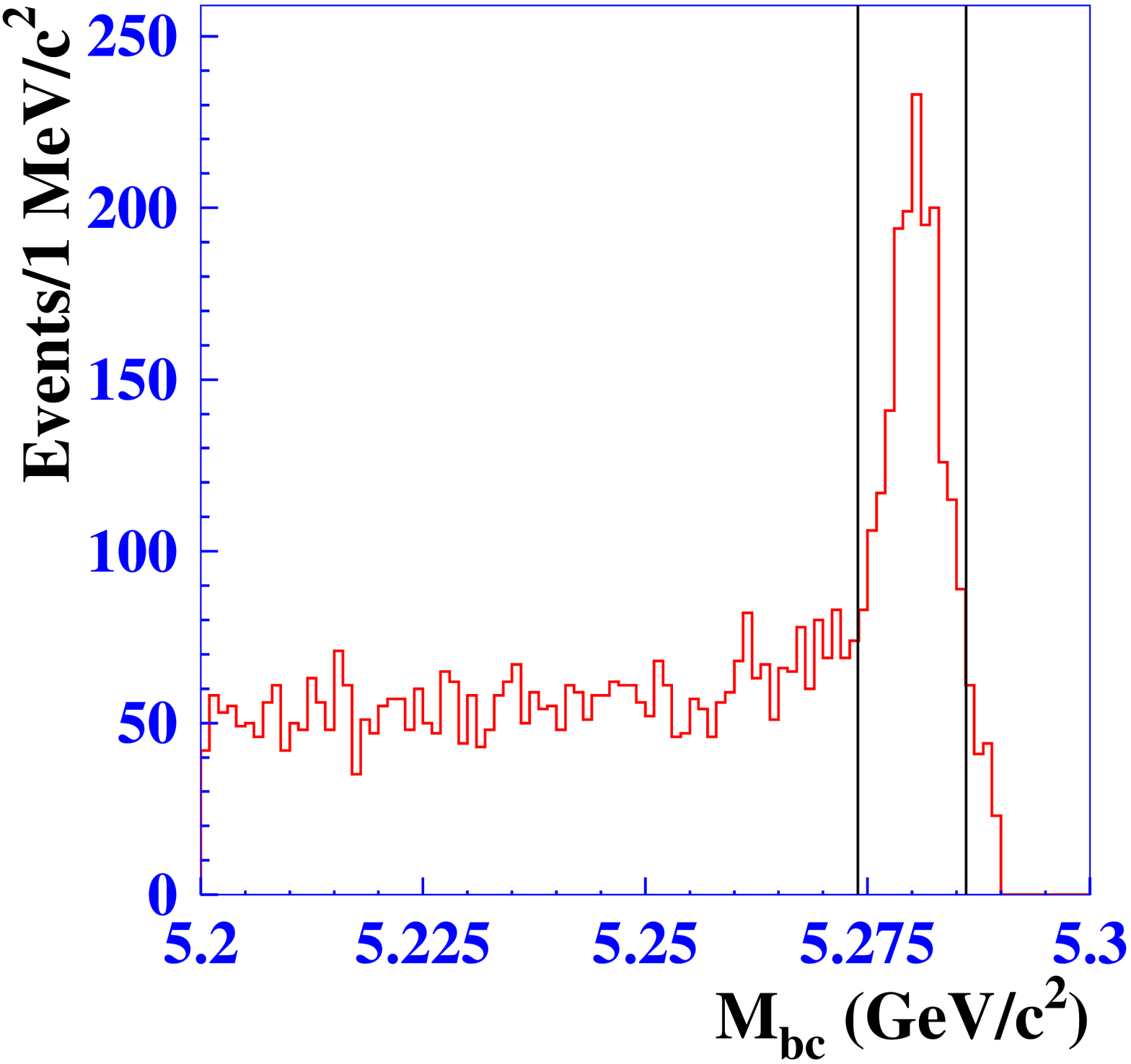}&
\includegraphics[height=6 cm]{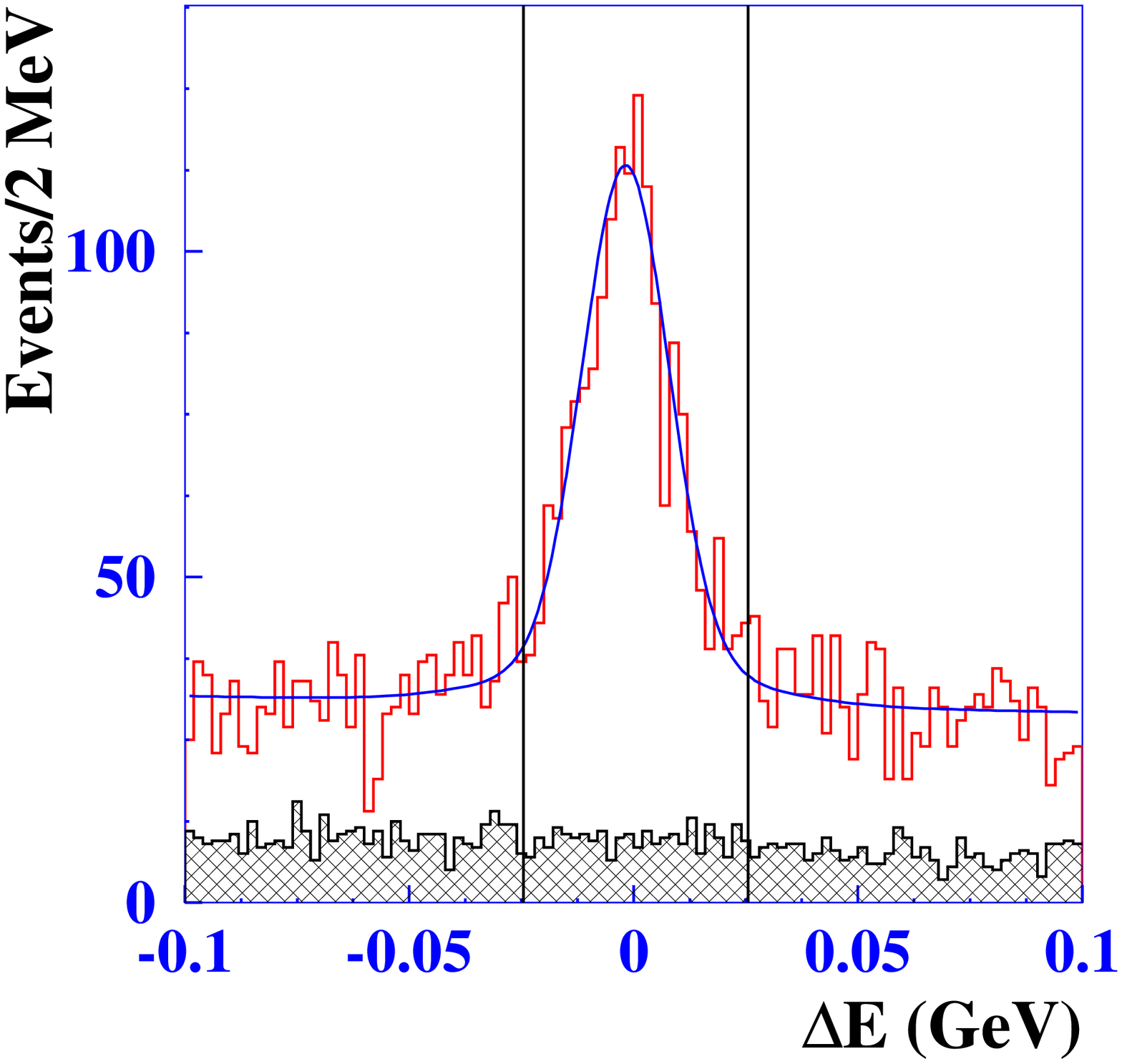}\\
\vspace*{-6.3 cm} & \\
{\hspace*{4.9cm}\bf\large a)}&{\hspace*{4.9cm}\bf\large b)}\\
\vspace*{4.8 cm} & \\
\end{tabular}
\caption{
The a) $M_{\rm bc}$ and
b)  $\Delta E$ distributions  for
 $B^{-}\to D^{+}\pi^{-}\pi^{-}$ events. The hatched histogram in (b) is 
the $D$ mass sideband 
($\big| |M_D-M_{K\pi\pi}|-26\,\rm MeV/c^2\big|<6.5\,\rm MeV/c^2$). 
 }
\label{f:dpmbde}
\end{figure}

 The $M_{\rm bc}$  and $\Delta E$ distributions
for $B^{-}\to D^{+}\pi^{-}\pi^{-}$ events 
are shown in Fig.~\ref{f:dpmbde}. 
The distributions are plotted for events that satisfy 
the selection criteria for the other variable: i.e.
$|\Delta E|<25$~MeV and $|M_{\rm bc}-M_B|<6$~MeV/c$^2$ for the 
$M_{\rm bc}$ and  $\Delta E$ histograms, respectively. A clear signal 
is evident in both distributions.
The signal yield is obtained  by fitting the  $\Delta E$ 
distribution to the  sum of  two Gaussians
with the same mean for the signal and a linear
 function for background. The widths and the relative normalization of the two Gaussians 
are fixed at values obtained from the MC simulation while 
the signal normalization  as well as the 
constant term and slope  of the background linear function
are treated as free parameters.

The signal yield  is $1101\pm46$ events.
The detection efficiency of $(18.2\pm 0.2)\%$  is determined from a MC
simulation that uses a Dalitz plot 
distribution that is 
 generated according to the model described 
in the next section.

Using the branching fraction 
${\cal {B}}(D^+\to K^-\pi^+\pi^+)=(9.1\pm0.6)\%$~\cite{PDG},
we obtain
$$
{\cal B}(B^-\to D^+\pi^-\pi^-)=(1.02\pm0.04\pm0.15)\times10^{-3},
$$
which is consistent with the upper limit obtained by CLEO,
$
{\cal B}(B^-\to D^+\pi^-\pi^-)<1.4\times 10^{-3}
$~\cite{CLEOd}. The statistical significance of the signal
is greater than 25$\sigma$~\cite{foot2}.
 This is the first observation of this decay mode.
The second error is systematic  and is dominated by a 10\% 
uncertainty in the track reconstruction (a $2\%$ per track uncertainty
was determined by comparing the signals for $\eta\to\pi^+\pi^-\pi^0$ 
and $\eta\to\gamma\gamma$).
 The uncertainty in 
the $D^+\to K^-\pi^+\pi^+$ branching fraction is 
$6.6\%$ and that for the  particle identification 
efficiency is $5\%$. Other contributions are 
smaller. The uncertainty in the background shape 
is estimated by adding higher order polynomial terms to the fitting
function,  which results in less than  a $5\%$ 
change in the branching fraction. 
The MC simulation uncertainty is estimated to be $3\%$.
The possible contribution from charmless $B$-meson decay modes is
estimated from the $M_D $ sidebands. The sideband distribution, 
shown as the hatched histogram in Fig.~\ref{f:dpmbde}(b), indicates
no excess from such events in the signal region.

\subsection{$B\to D\pi\pi$ Dalitz plot analysis}
For a three-body decay of a spin zero particle, two 
variables are required to describe the decay kinematics; 
we use the two 
$D\pi$ invariant masses.
Since there are two 
identical pions in the final state, we separate the pairs with maximal and 
minimal $M_{D\pi}$ values.

\begin{figure}[h]
\begin{center}
\begin{tabular}{cc}
\includegraphics[width=8 cm]{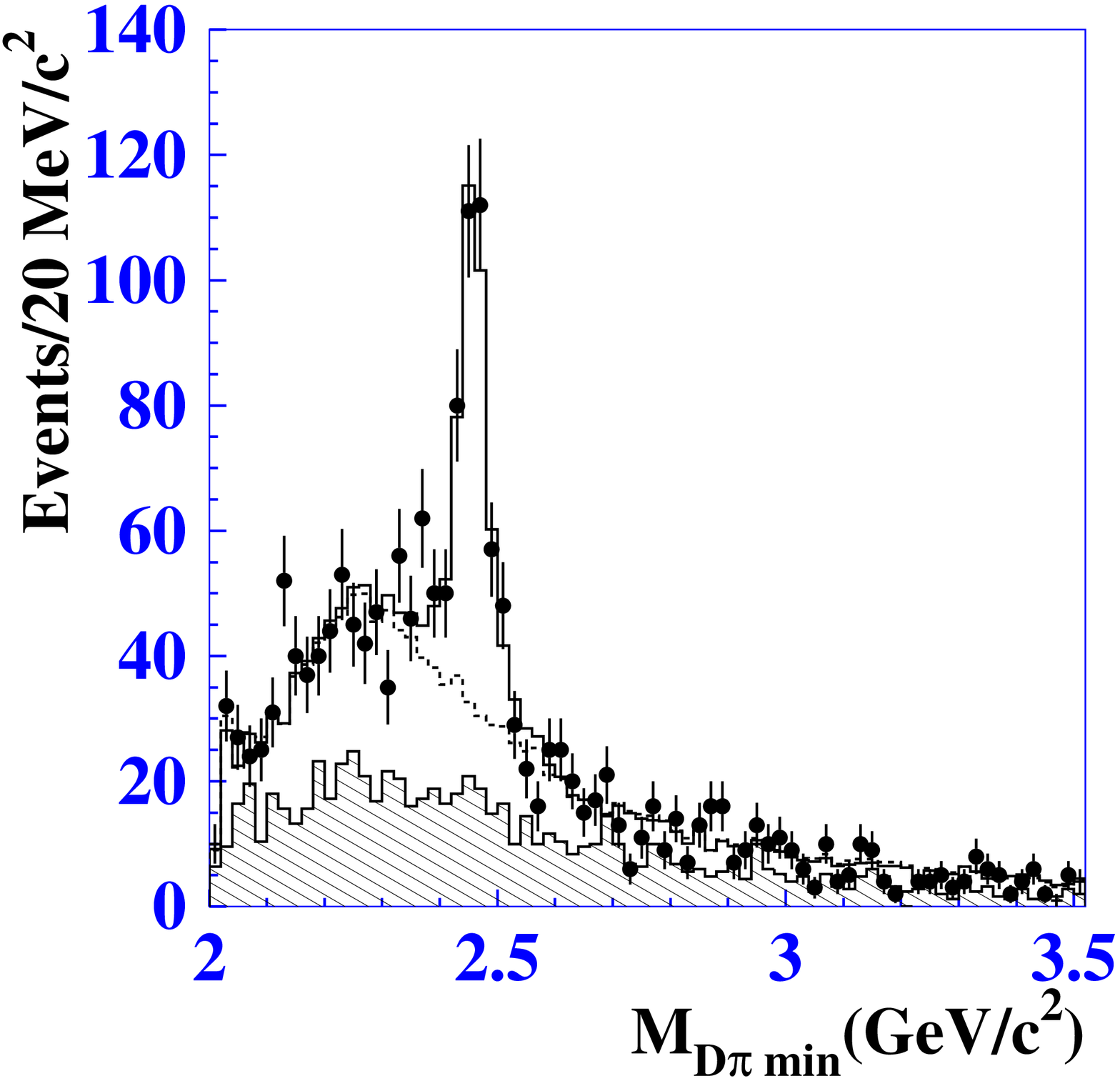}&
\includegraphics[width=8 cm]{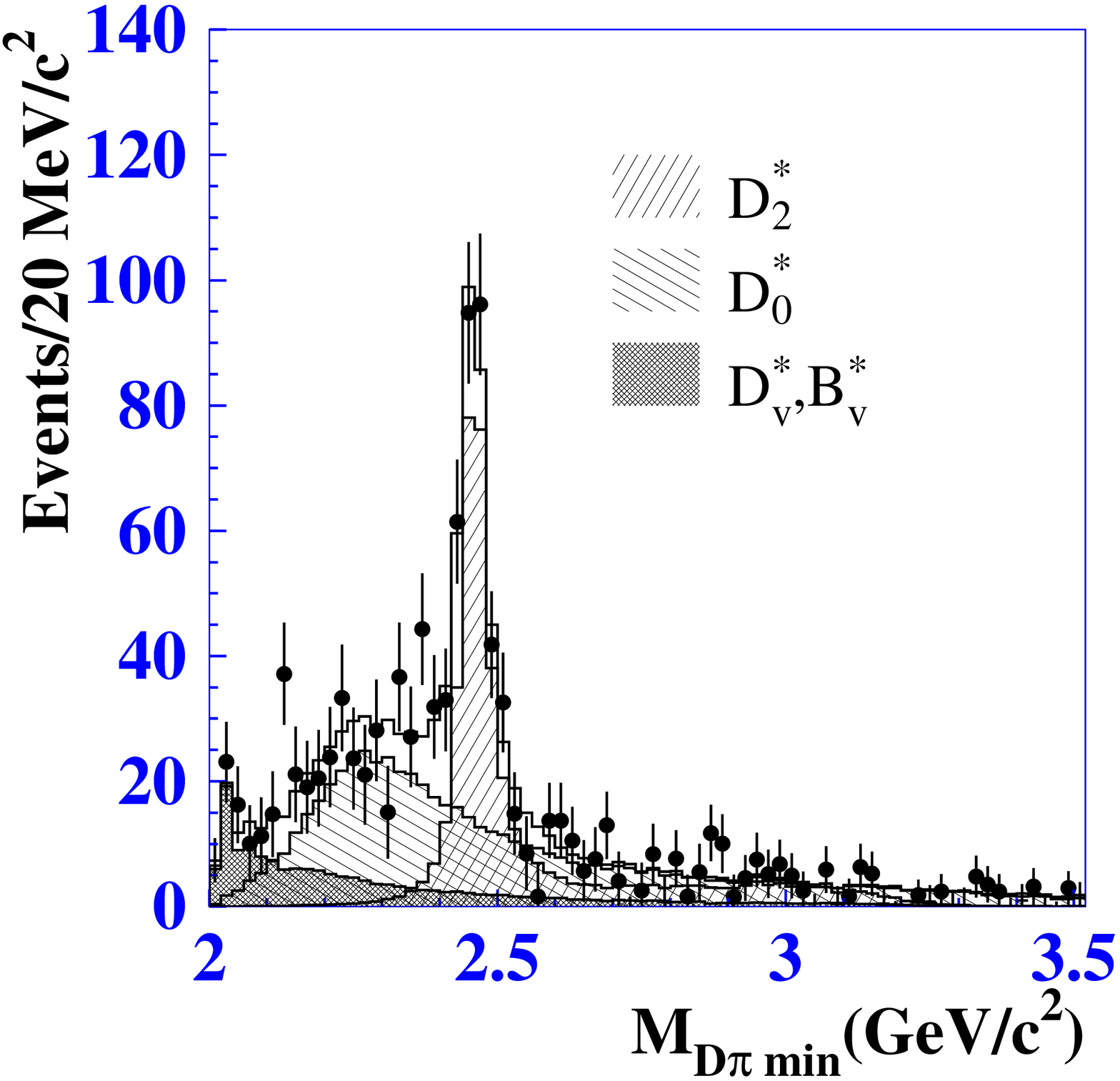}\\
\vspace*{-8.1 cm} & \\
{\hspace*{6.7cm}\bf\Large a)}&{\hspace*{6.7cm}\bf\Large b)}\\
\vspace*{6.6 cm} & \\
\end{tabular}

\caption{a)The  minimal $D\pi$ mass distribution of 
$B^-\to D^+\pi^-\pi^-$ candidates. 
The points with error bars correspond to 
the signal box events,  while the hatched histogram
shows the background obtained from the sidebands. 
The open histogram is the result of a 
fit while the dashed one shows the fit function  in the case when the  
narrow resonance amplitude is set to zero. b)The background-subtracted $D\pi$ mass distribution. The points  with error bars correspond to 
the signal box events, hatched histograms show different
contributions, the open histogram shows 
the coherent sum of all contributions.  }
\label{f:dpp_M}
\end{center}
\end{figure}

To analyze  the dynamics of $B\to D\pi\pi$ decays,  events 
with $\Delta E$ and $M_{\rm bc}$
within the 
$|\Delta E|<25$~MeV,  $|M_{\rm bc}-M_B|<6~{\rm MeV}/c^2$ signal region
are selected.
To model  the contribution and shape of the background, we use
a sideband region defined as  100~MeV~$>|\Delta E|>$~30~MeV with 
the signal $M_{\rm bc}$ given above.
The minimal $D\pi$ mass distributions for the signal and sideband events 
are shown in Fig.~\ref{f:dpp_M}, where narrow and broad resonances are 
visible.

\begin{figure}[h]
\begin{center}
\begin{tabular}{cc}
\includegraphics[width=8 cm]{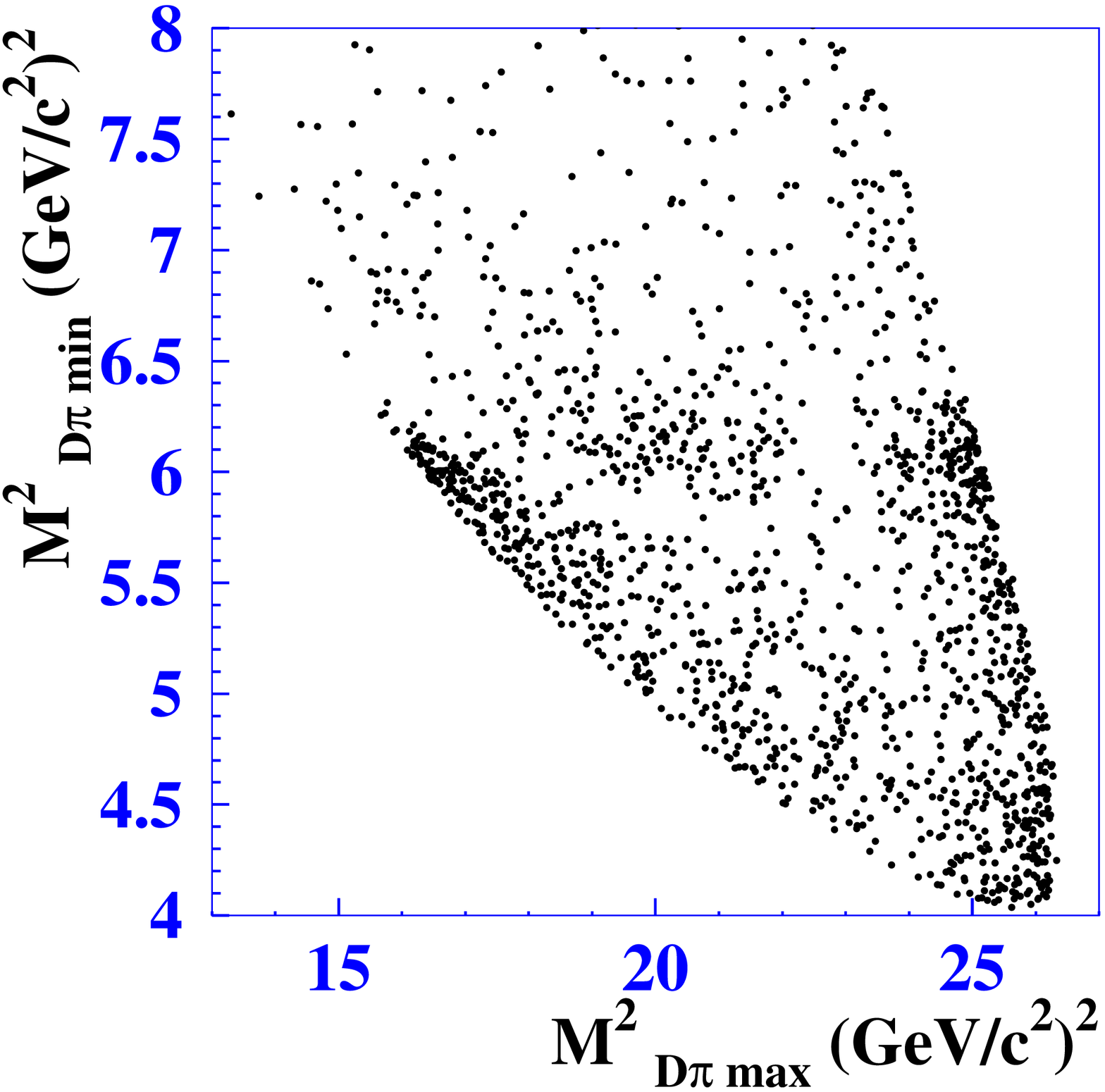}&
\includegraphics[width=8 cm]{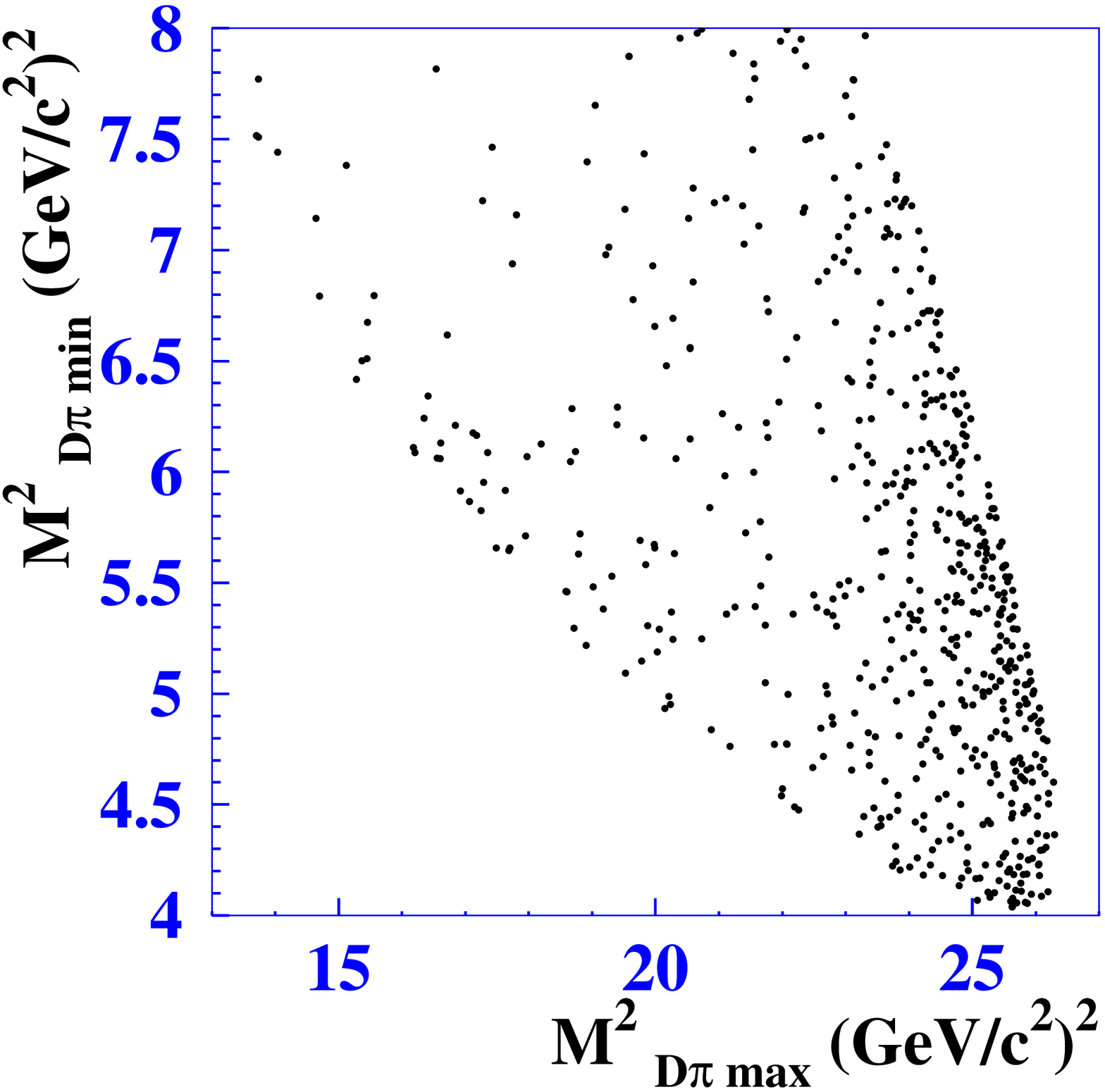}\\
\vspace*{-8.1 cm} & \\
{\hspace*{6.7cm}\bf\Large a)}&{\hspace*{6.7cm}\bf\Large b)}\\
\vspace*{6.6 cm} & \\
\end{tabular}
\caption{The Dalitz plot for a) signal events and  b) sideband events.}
\label{f:dpp_DP}
\end{center}
\end{figure}

The distributions of events in the $M^2_{D\pi~min}$ 
versus $M^2_{D\pi~max}$
Dalitz plot for the signal and sideband regions
are shown 
in Fig.~\ref{f:dpp_DP}. 
The Dalitz plot boundary is determined by the
decay kinematics and the masses of their daughter particles. 
In order to have the same Dalitz plot boundary for events  
in both signal and sideband regions,
mass-constrained fits 
of 
$K\pi\pi$ to $M_D$ and $D\pi\pi$ to 
$M_B$ are performed. 
The mass-constrained fits also reduces the smearing from
detector resolution. 

To extract the amplitudes and phases of different intermediate states,
an   unbinned fit to the Dalitz plot is performed using a method
similar 
to CLEO's~\cite{CLEODP}.
The event density function in the Dalitz plot is the sum of the signal  
and  background.

Since the  $D\pi$ mass distributions for the upper  
and lower halves of the $\Delta E$  sideband have similar shapes, we
can expect similar background behavior for the signal and sideband regions.
The background shape
is obtained from an unbinned fit of 
the sideband distribution to a smooth two-dimensional function.
The number of background events in the signal region 
is scaled according to the relative areas of the signal and the 
sideband regions. 

In the $D^+\pi^-\pi^-$ final state 
a  combination of the  $D^+$-meson and a pion can form either a tensor meson 
$D^{*0}_2$ or a scalar state $D_0^{*0}$; the axial vector mesons $D_1^0$ and 
$D'^{0}_1$ cannot decay to two pseudoscalars because of the angular 
momentum and parity conservation.
The $D^{*0}$ cannot decay to the $D^{+}\pi^-$ because  the 
$D^{*0}$ mass  is lower than that of $D^{+}\pi^-$. However,  a
virtual $D^*_v$ of a higher mass can decay to 
$D^{+}\pi^-$.
Another virtual hadron that can be produced in this combination is $B^{*}_v$
($B\to B^*_v\pi$ and $B^*_v\to D\pi$). The contributions of these intermediate states 
are included in the signal-event density ($S(q_1^2,q_2^2)$)
parameterization 
as a coherent sum of the 
corresponding amplitudes
together with a possible constant amplitude ($a_3$):
\begin{equation}
\label{e:dpp}
S(q_1^2,q_2^2)=\left|a_{D^{*}_2}A^{(2)}(q_1^2,q_2^2)+a_{D^{*}_0}e^{i\phi_{D^{*}_0}}A^{(0)}
(q_1^2,q_2^2)+a_{D^*_v}e^{i\phi_{D^*_v}}A^{(1)}(q_1^2,q_2^2)
+a_{B^*_v}e^{i\phi_{B^*_v}}A^B(q_1^2,q_2^2)+a_3e^{i\phi_3}\right|^2\otimes {\cal{R}}(\Delta q^2),
\end{equation}
where $\otimes {\cal{R}}(\Delta q^2)$ denotes convolution with the experimental resolution.
Each resonance is described by a relativistic Breit-Wigner with a
$q^2$ dependent width 
and an angular dependence that corresponds to the spins of the
intermediate and final state particles:
\begin{equation}
\label{e:brd}
A^{(L)}(q_1^2,q_2^2) = F_{BD^{**}}^{(L)}
(\mathbf{p_{1}})\frac{T^{(L)}(q_1,q_2)}
{q_1^2-M_L^2+iM_L\Gamma_L(q_1^2)}F_{D^{**}D}^{(L)}(\mathbf{p_{2}})+(q_1 \to q_2),
\end{equation}
where 
\begin{equation}
\label{gmd}
\Gamma_L(q^2)=\Gamma_L\cdot (\mathbf{p_{2}/p_{2}^0})^{2L+1}(M_L/\sqrt{q^2})F_{D^{**}D}^{(L)2}(\mathbf{p_{2}})
\end{equation}
is the $q^2$ dependent width of the $D^{**}$, 
with mass $M_L$ and width $\Gamma_L$, decaying to $D\pi$ state with
orbital angular momentum $L$.
The variables $p_1,~p_2,~p_D,~q_1=p_2+p_D,~q_2=p_1+p_D$ are
the four-momenta of the pions,
$D$, and $D\pi$ combinations, respectively; 
${\mathbf{ p_{2},~p_{2}^0}}$  are the magnitude of 
the pion three-momentum  in 
the $D^{**}$ rest frame when the $D^{**}$ has a four-momentum-square equal 
to $q^2$ and $M_L^2$, respectively;
${\mathbf{ p_{1},~p_{1}^0}}$  are the magnitude of the pion 
three-momentum  in 
the $B$ rest frame for the case when the  $D^{**}$ four-momentum squared
is  equal to $q^2$ and $M_L^2$, respectively.
The angular dependence for different spins of the intermediate 
states is:   
\begin{equation}
\label{eqnm}
T^{(0)}(q_1,q_2)=1,~
T^{(1)}(q_1,q_2)=\frac{M_B\mathbf{p_{2}p_{1}}}{\sqrt{q_1^2}}\cos\theta,~
T^{(2)}(q_1,q_2)=\frac{M_B^2\mathbf{p_{2}^2p_{1}^2}}{{q_1^2}}(\cos^2\theta-1/3),
\end{equation}
where $\theta$  is the angle between the first pion from the $B$-decay
and the pion from the $D^{**}$-decay in 
the $D^{**}$ rest frame, and
$ F_{BD^{**}}^{(l)}(\mathbf{p_{1}})$ and  $F_{D^{**}D}^{(l)}(\mathbf{p_{2})}$ are transition
form factors, which are the most uncertain part of the resonance description. 
For the $B\to D^{**}$ and $D^{**}\to D$ form factors,  we use  
the Blatt-Weiskopf parameterization~\cite{blat}:
\begin{equation}
\label{eqfo}
F^{(0)}_{AB}(\mathbf{p})=1,~
F^{(1)}_{AB}(\mathbf{p})=\sqrt{\frac{1+(\mathbf{p^0}r)^2}{1+(\mathbf{p}r)^2}},~
F^{(2)}_{AB}(\mathbf{p})=\sqrt{\frac{9+3(\mathbf{p^0}r)^2+(\mathbf{p^0}r)^4}{9+3(\mathbf{p}r)^2+(\mathbf{p}r)^4}},
\end{equation}
where $r$=1.6~$\rm(GeV/c)^{-1}$ is a hadron scale.
For the virtual mesons $D^*_v$ and $B^*_v$ that are produced beyond the
peak region,
another form factor parameterization has been used: 
\begin{equation}
\label{eqf1}
F_{AB}(\mathbf{p})=e^{-r(\mathbf{p-p_0})};
\end{equation}
this provides stronger suppression of the numerator in
Eq.~(\ref{e:brd}) far from the
resonance region. 
The resolution function is obtained from MC simulation; 
the detector resolution for the $D\pi$ invariant mass   is about $4$~MeV/$c^2$.

The $D^{**}$ resonance parameters~($M_{D^{*0}_2},~\Gamma_{D^{*0}_2},~M_{D^{*0}_0},~\Gamma_{D^{*0}_0}$) as 
well as the amplitudes for the intermediate states and relative
phases~($a_{D^{*}_2},~a_{D^*_v},~a_{D^{*}_0},~a_{B^*},~a_3,~\phi_{D^*_v},~\phi_{D^{*}_0},~\phi_{B^*},~\phi_3$) 
are treated as free parameters in the fit. 

Table~\ref{t:dpp} gives the results of the fit for different models. 
When the $D_v^*$ amplitude is included, the likelihood  significantly improves
and gives 
branching fractions values  that are 
consistent with expectation based on the 
$D^*$ width and the $B^-\to D^{*0}\pi^-$ 
branching fraction. 
When the $B^*_v$ amplitude is added, the likelihood is also significantly
improved. A constant phase space term, $a_3 \exp(i\phi_3)$,
 does not substantially change  
the likelihood and the final results are presented without this term. 
The variation of the fit parameters when these last  
amplitudes are included 
is used as an estimate of the model error. 

\begin{table}
\begin{tabular}{|c|c|c|c|c|c|c|}
\hline
      & I      &II&& III && IV\\

Parameters                        & $D^*_2,~D^*_0$      & $D^*_2,~D^*_0,D^*_v$&& $D^*_2,~D^*_0,D^*_v,~B^*_v$ && $D^*_2,~D^*_0,D^*_v,~B^*_v,$\\
                      & &&&  && ph.sp$(a_3)$\\
\hline                                                                                                                                                                                                                         
\hline                                                                                                                                                                                                                         
$Br_{D_2^{*}}(10^{-4})$           &   3.21 $\pm$  0.24  &   3.26$\pm$  0.26   &&   3.38 $\pm$  0.31   &&   3.47 $\pm$  0.37        \\
$Br_{D_0^{*}}(10^{-4})$           &   6.09 $\pm$  0.42  &   4.96 $\pm$  0.47  &&   6.12 $\pm$  0.57   &&   8.35 $\pm$  0.94        \\
$\phi_{D_0^{*}}$                  &  -2.01 $\pm$  0.10  &  -2.35 $\pm$  0.11  &&  -2.37 $\pm$  0.11   &&  -2.31 $\pm$  0.14        \\
$Br_{D^*_v}(10^{-4})$             &   --  &   1.46 $\pm$  0.23  &&   2.21 $\pm$  0.27   &&   2.23 $\pm$  0.32        \\
$\phi_{D^*_v}$                    &      --             &   0.03 $\pm$  0.15  &&  -0.25 $\pm$  0.15   &&  -0.33 $\pm$  0.19        \\
$Br_{B^*_v}(10^{-4})$             &      --             &    --
&&   0.67 $\pm$  0.04   &&   0.72 $\pm$  0.04        \\
$\phi_{B^*_v}$                      &      --             &  --
&&  -0.27 $\pm$  0.28   &&  -0.39 $\pm$  0.24        \\
$M_{D_2^{*0}}(MeV/c^2)$           & 2454.6 $\pm$   2.1  & 2458.9 $\pm$   2.1  && 2461.6 $\pm$   2.1   && 2462.7 $\pm$   2.2        \\
$\Gamma_{D_2^{*0}}(MeV)$          &   43.8 $\pm$   4.0  &   44.2 $\pm$   4.1  &&   45.6 $\pm$   4.4   &&   46.1 $\pm$   4.5        \\
$M_{D_0^{*0}}(MeV/c^2)$           &  2268  $\pm$   18   &  2280  $\pm$   19   &&  2308  $\pm$   17    &&  2326  $\pm$   19         \\
$\Gamma_{D_0^{*0}}(MeV)$          &   324  $\pm$   26   &   281  $\pm$
23   &&   276  $\pm$   21    &&   333  $\pm$   37       \\
$a_3\times 10^5 $          &  -- &   --   &&    --  && $0.38   \pm
0.65$        \\
$\phi_3 $          &  -- &   --   &&    --  && $-0.10  \pm 0.93$        \\
\hline                                                                                                                
$N_{sig}$                         &  1058  $\pm$   47   &  1007  $\pm$   44   &&  1056  $\pm$   46    &&  1068  $\pm$   47         \\
\hline                                                                                                                
$-2\ln\cal{L/L_0}$                &        115         &       26
                      &&    0           &&     -7              \\
$\chi^2/N$                        &    253.9/129        &    185.2/127        &&   166.5/125          &&    158.5/123              \\
\hline
\end{tabular}
\caption{Fit results for different models.
The model used to obtain the results includes amplitudes for
$D^*_2,~D^*_0,D^*_v,~B^*_v$ intermediate resonances. Adding the constant
 term (ph.sp$(a_3)$) does not  significantly improve the likelihood.
}
\label{t:dpp}
\end{table}

\begin{figure}[h]
\begin{center}
\includegraphics[width=10 cm]{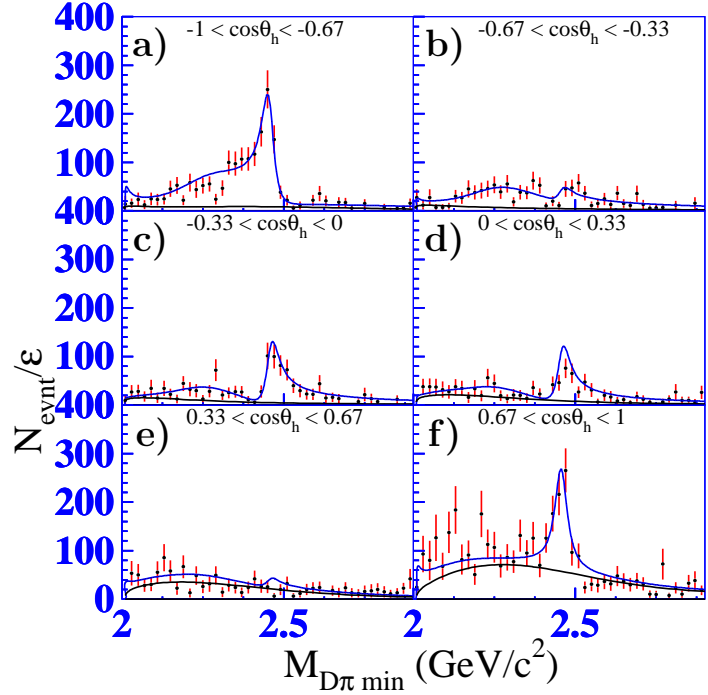}
\end{center}

\vspace*{-9.8 cm} 
{\hspace*{-1.8cm}\bf\Large a)}{\hspace*{3.4cm}\bf\Large b)}

\vspace*{2.05 cm} 
{\hspace*{-1.8cm}\bf\Large c)}{\hspace*{3.4cm}\bf\Large d)}

\vspace*{2.05 cm}
{\hspace*{-1.8cm}\bf\Large e)}{\hspace*{3.4cm}\bf\Large f)}

\vspace*{2.9 cm}
\caption{The minimal $D\pi$ mass distribution for different helicity ranges. 
The two curves are the fit results
for the case of $D^*_2,~D_0^*$ and $D^*_v$ amplitudes (the top curve) and the
background contribution (the bottom one). 
The number of events in each bin is corrected for the efficiency 
obtained from MC simulation.}

\label{f:dpp_mdpf}
\end{figure}

Figure~\ref{f:dpp_mdpf} shows the $D\pi$ mass distributions for
different $D\pi$ helicity regions. 
The helicity ($\cos{\theta_h}$) is defined as the cosine of the 
angle between the pions from the  $B$ and $D^{**}$ decays in 
the rest frame of $D^{**}$.
The number of events in each bin is corrected for the MC-determined 
efficiency. 
The curve shows the fit function 
for the case when the $D^*_2,~D_0^*$, $D^*_v$ and $B^*_v$ 
amplitudes are included.
The $D^*_2$ resonance is clearly 
seen in  the helicity range  $|\cos\theta_h|>0.67$, 
where the D-wave peaks.  
The range  $0.33<|\cos\theta_h|<0.67$ where the D-wave amplitude is 
suppressed,  shows the 
S-wave contribution  from the $D_0^*$ while the low helicity 
range $|\cos\theta_h|<0.33$ demonstrates a clear interference pattern. 

\begin{figure}[h]
\begin{center}
\begin{tabular}[c]{cc}
\includegraphics[height=6 cm]{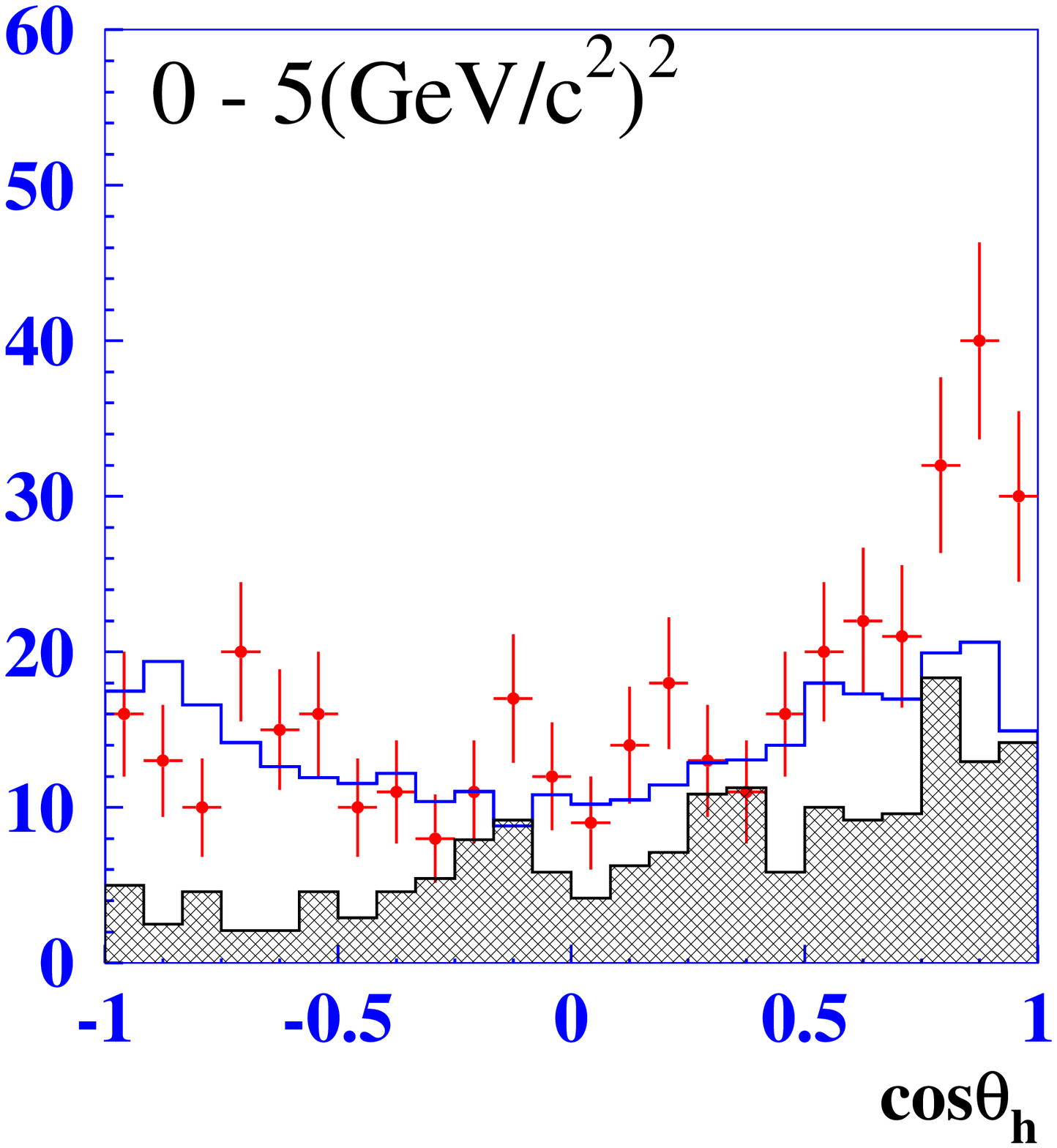}&
\includegraphics[height=6 cm]{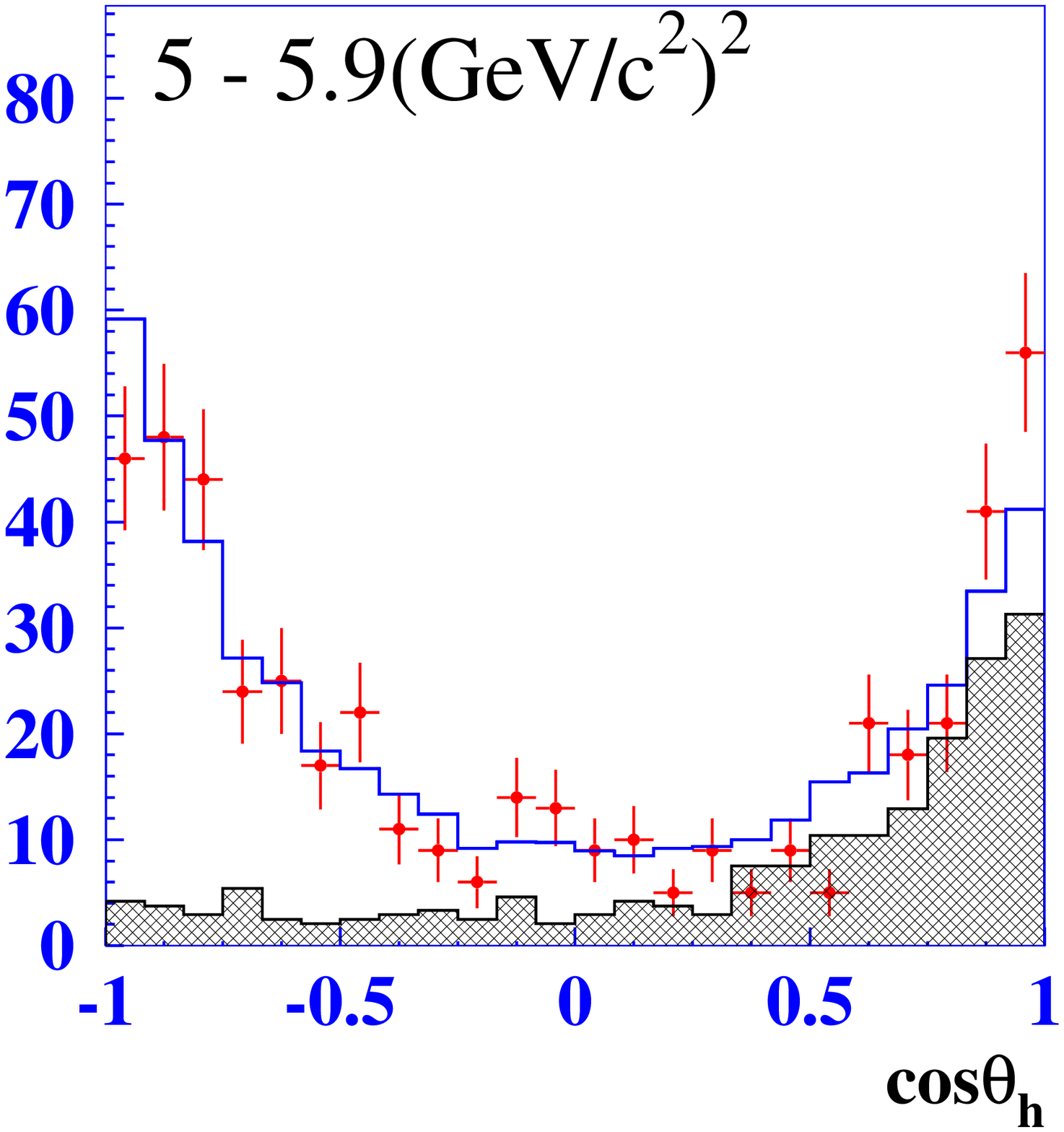}\\
\includegraphics[height=6 cm]{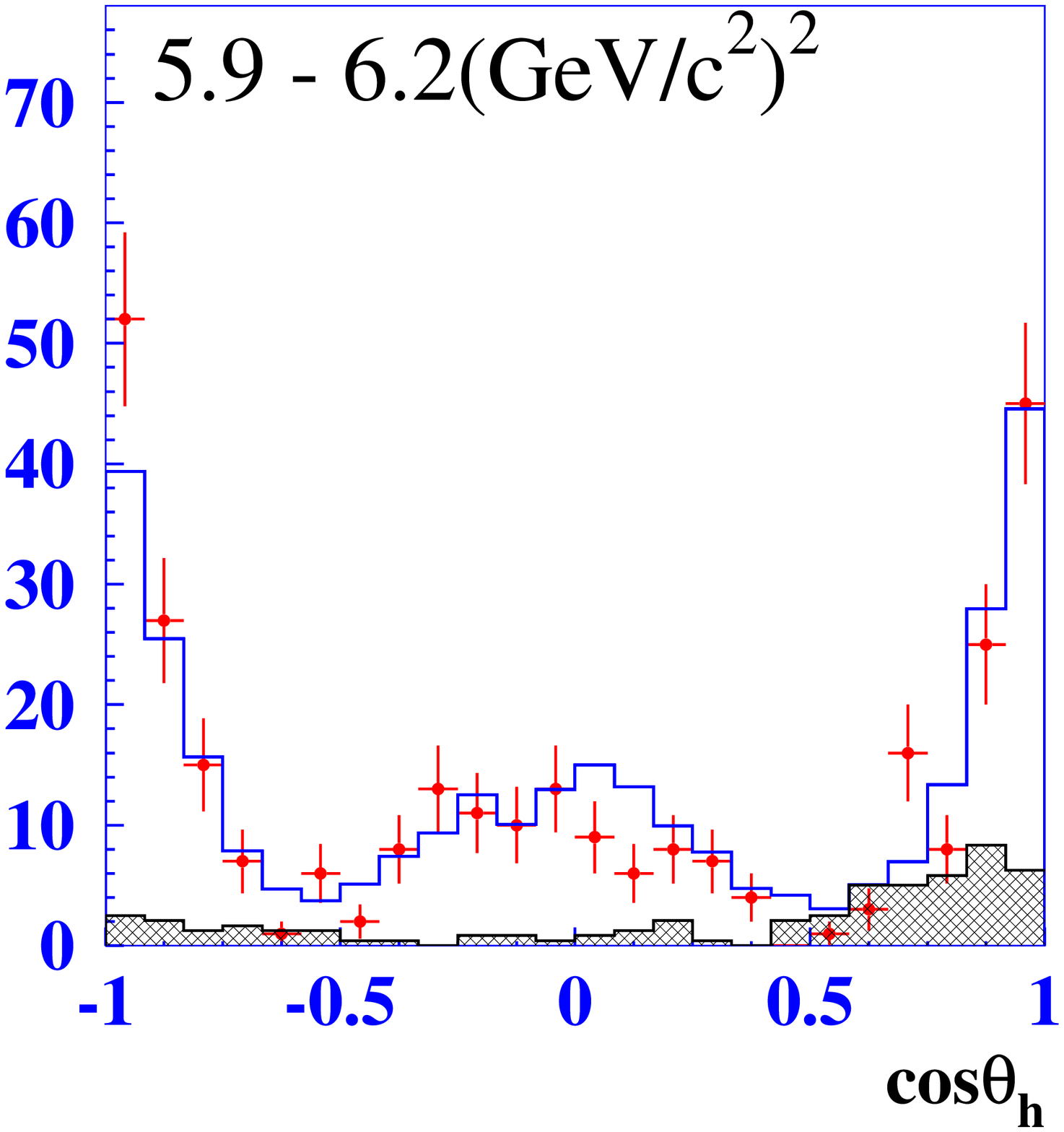}&
\includegraphics[height=6 cm]{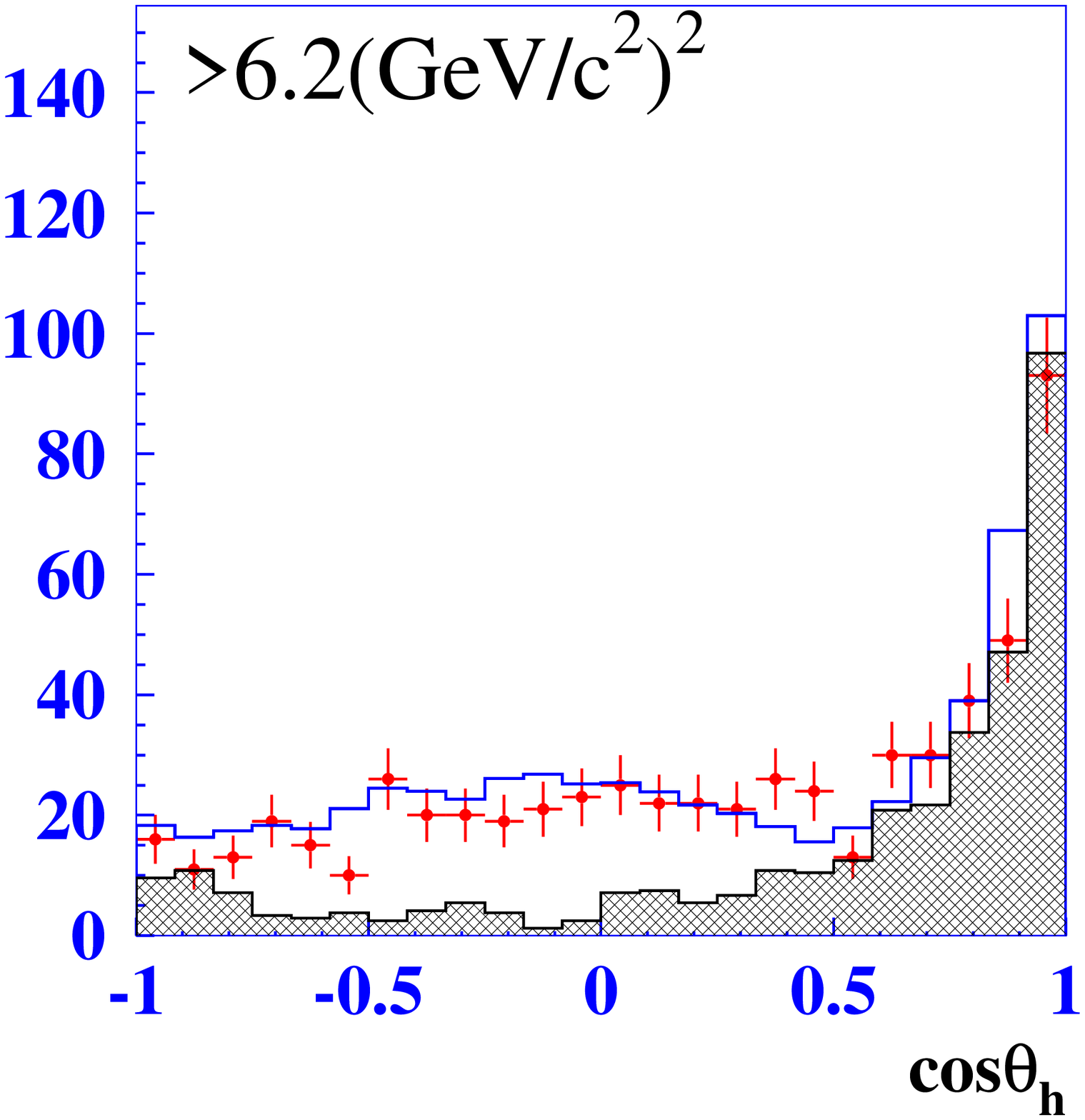}\\
\vspace*{-12.5 cm} & \\
{\hspace*{5.cm}\bf\Large a)}&{\hspace*{5.cm}\bf\Large b)}\\
\vspace*{5.1 cm} & \\
{\hspace*{5.cm}\bf\Large c)}&{\hspace*{5.cm}\bf\Large d)}\\
\vspace*{4. cm} & \\
\end{tabular}
\end{center}
\caption{The helicity distribution for data (points with error bars) and for 
MC simulation (open histogram). The hatched distribution shows 
the scaled background distribution from the $\Delta E$ sideband region.
Figures a), b)  correspond to the  $q^2$ region below $D_2^0$ resonance,
c) -- region of the tensor resonance, d) -- region higher of the $D_2^0$.
}
\label{f:dpp_hel}
\end{figure}

Another demonstration of the agreement between the fitting function and 
the data is given in Fig.~\ref{f:dpp_hel}, where the helicity 
distributions for different  $q^2$ regions are shown. The histogram 
in the region of the $D^*_2$ meson clearly indicates a 
$|\cos^2{\theta_h}-1/3|^2$ D-wave dependence. 
The 
distributions in the other  regions show  reasonable agreement between
the fitting function and the data except for a few bins in the  
small $M_{D\pi~min}$ region and  with
helicity close to 1 (Fig.~\ref{f:dpp_hel}(a)). 
This region is populated mainly by the virtual $D^*_v$ and $B^*_v$ 
production, the description of which depends on
the form factor behavior. 
This discrepancy does
not affect the determination of the $D^{**}$ parameters
that are the main topic of this work. 

The fit
quality is estimated using a two-dimensional histogram of minimum
$q^2_{D\pi}$ versus the  $D\pi$  helicity and calculating the
$\chi^2/N$ for the  function obtained from unbinned likelihood minimization.
The confidence level 
for the model with $D^*_2,~D^*_0,~D^*_v$ and $B^*_v$ is about $0.8\%$. 
The low confidence level is due to the poor description in the region
where $M_{D\pi~min}$ is small and $M_{D\pi~max }$ is
large (or helicity is close to 1) as discussed above.

In Table~\ref{t:d0v}, the likelihood 
values are presented for the case when the broad scalar resonance is excluded 
or when it has quantum numbers different from $J^P=0^+$. 
For all cases the likelihood values are  significantly worse. 
\begin{table}
\begin{center}
\begin{tabular}{|c|c|}
\hline
Model & $-2\ln({\cal L}/{\cal L}_{max})$   \\
\hline
\hline
${D^{\star}_2,~D^{\star}_0,~D^{\star}_v,~B^{\star}_v}$ & 0\\
\hline
${D^{\star}_2,~D^{\star}_v,~B^{\star}_v,}$ ph.sp$(a_3)$ & 265 \\
${D^{\star}_2,~D^{\star}_v~,~B^{\star}_v,1^-}$ & 355 \\
${D^{\star}_2,~D^{\star}_v~,~B^{\star}_v,2^+}$ & 235\\
\hline
\end{tabular}
\caption{Comparison of  models with and without a $0^+$ resonance.
The amplitudes for $D_2^*$ and the virtual $D_v^*$ and  $B_v^*$ are always included. }
\label{t:d0v}
\end{center}
\end{table}
Thus, we claim  the observation of  a broad  state that can be 
interpreted as the scalar $D^*_0$.
The fit gives the following parameter values:
$$
M_{D^{*0}_0}=(2308\pm17\pm15\pm28) {\rm MeV}/c^2,~~\Gamma_{D^{*0}_0}=(276\pm21\pm18\pm60)\rm{MeV}.
$$
The values corresponds to the case when four
amplitudes (column III of Table~\ref{t:dpp}) were 
included.
Here and throughout the paper the first error is statistical, the second 
is systematic  and the third is the model-dependent error 
described below.

The values of the narrow resonance mass and width obtained from the fit are:
$$
M_{D^{*0}_2}=(2461.6\pm2.1\pm0.5\pm3.3) {\rm MeV}/c^2,~~\Gamma_{D^{*0}_2}=(45.6\pm4.4\pm6.5\pm1.6){\rm MeV}.
$$
The value of the $D^{*0}_2$ width is larger than the world  average of 
$23\pm5$~MeV, and is consistent with the preliminary result from  FOCUS of 
$30.5\pm4.2$~MeV~\cite{e831}. 
The previous analyses did not take the interference of
intermediate states into account; this suggests that there
may be large unaccounted systematic errors in these
measurements.

The following branching ratio products  are obtained:
{
$$
{\cal B}(B^-\to D^{*0}_{2}\pi^-)\times {\cal B}(D_2^{*0}\to D^{+}\pi^-)=(3.4\pm0.3\pm0.6\pm0.4)\times10^{-4},
$$
$$
{\cal B}(B^-\to D^{*0}_0\pi^-)\times {\cal B}(D_0^{*0}\to D^{+}\pi^-)=(6.1\pm0.6\pm0.9\pm1.6)\times10^{-4},
$$
and the relative phase of the scalar and tensor amplitude is
$$
\phi_{D^{*0}_0}=-2.37\pm 0.11\pm0.08\pm0.10.
$$
}
The systematic errors are estimated by comparing
the fit results for the case when the  background shape is taken 
separately from the lower or upper 
sideband in the $\Delta E$ distribution. The fit is also performed with more 
restrictive cuts on $\Delta E$, $M_{\rm bc}$ and $\Delta M_D$. 
The maximum difference is taken as an additional estimate of the systematic
uncertainty.   
For branching fractions, the systematic errors also include 
uncertainties in  track 
reconstruction and PID efficiency,
as well as the error in the $D^+\to K^-\pi^+\pi^+$ absolute branching fraction.

The model uncertainties are estimated by comparing fit results for the case 
of different models (II-IV in Table~\ref{t:dpp}) and for values of $r$
that range from 0 to 5 (GeV/c)$^{-1}$ for  the transition form factor 
defined in Eqs.~(\ref{eqfo}) and (\ref{eqf1}).

\section{$B^{-}\to D^{*+}\pi^{-}\pi^{-}$ analysis}

\begin{figure}[h]
\begin{tabular}{cc}
\includegraphics[height=6 cm]{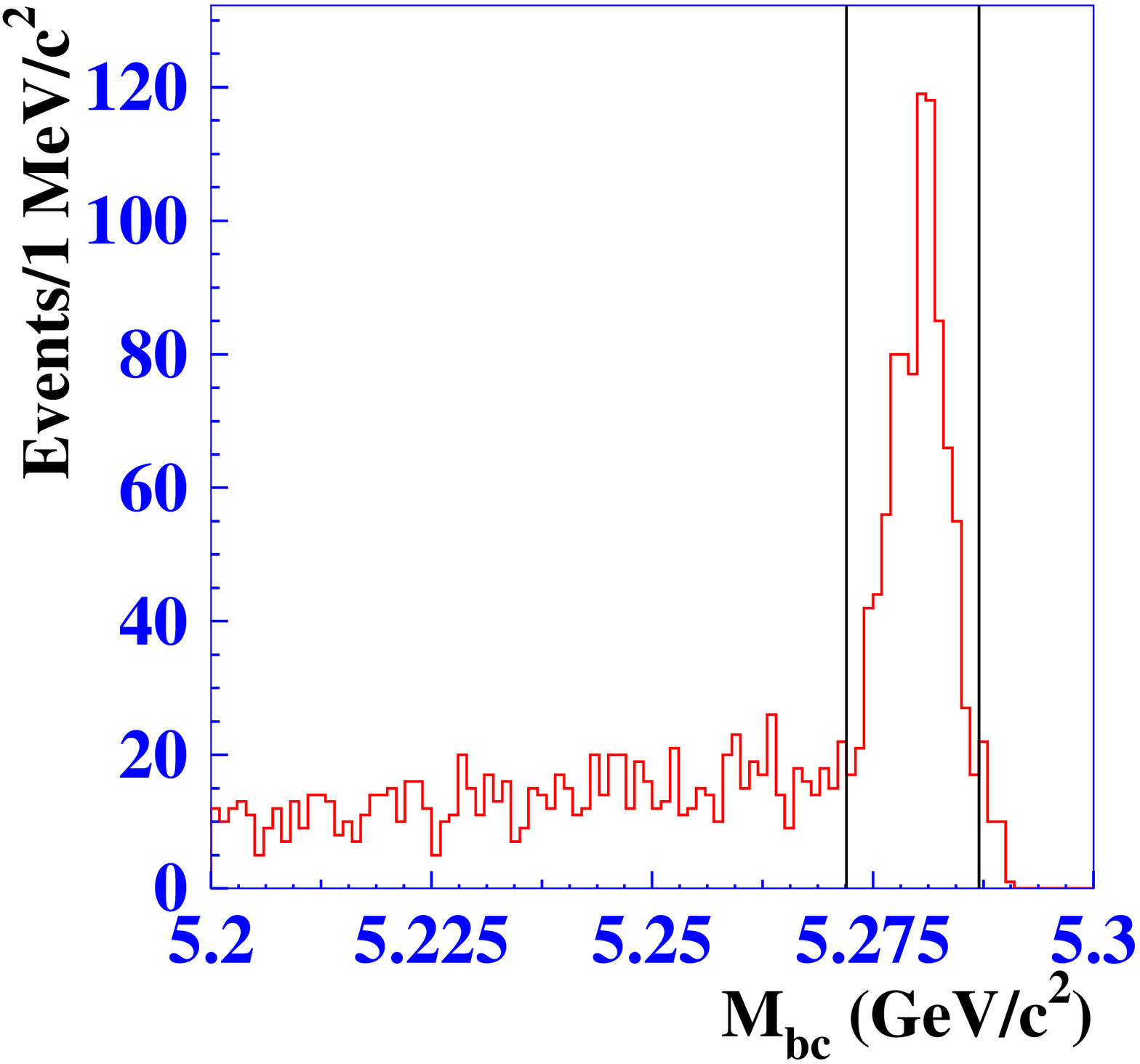}&
\includegraphics[height=6 cm]{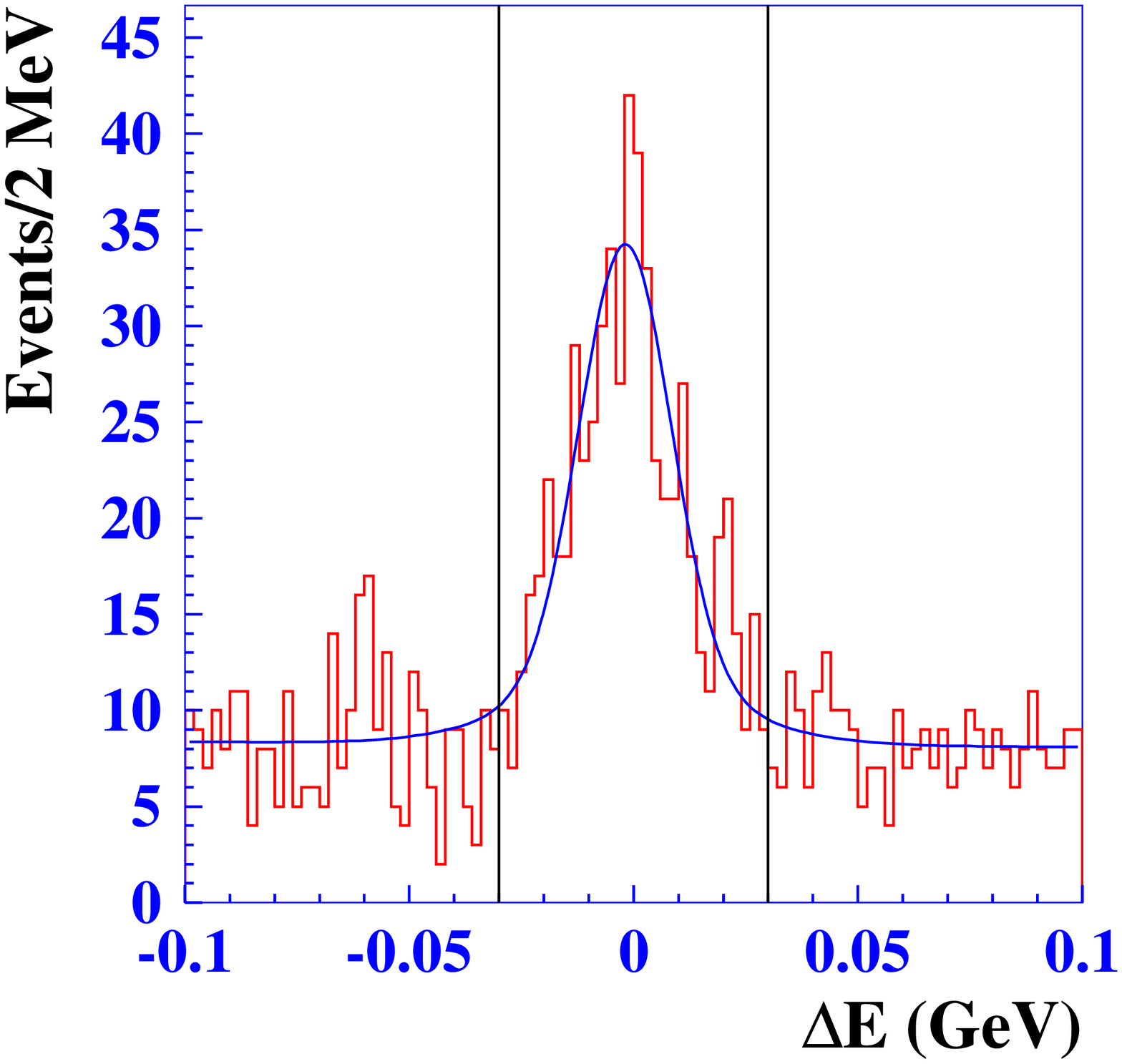}\\
\vspace*{-6.3 cm} & \\
{ \hspace*{4.8cm}\bf\Large  a)}&{\bf\Large  \hspace*{4.7cm} b)}\\
\vspace*{4.2 cm} & \\
\end{tabular}
\caption{
The a) $M_{\rm bc}$  and
b) $\Delta E$ distributions 
for $B^{\mp}\to D^{*\pm}\pi^{\mp}\pi^{\mp}$ candidates.
}
\label{f:dsmbde}
\end{figure}

For $D^*$ reconstruction,  the $D^{*+}\to D^0\pi^+$ decay is used and two  
decay modes  $D^0\to K^-\pi^+$ and 
$D^0\to K^-\pi^+\pi^+\pi^-$  are included.
The $\Delta E$ and $M_{\rm bc}$ distributions  
are shown in Fig.~\ref{f:dsmbde}.
In each mode the number of signal events is obtained in a way  similar to 
that described for the $D\pi\pi$ selection. 
The observed signal yields of
$N_{K\pi}=273\pm21$ and $N_{K3\pi}=287\pm22$ for the $K\pi$ and
$K\pi\pi\pi$ modes,
respectively, are consistent, based on the $D$
branching fractions and the
efficiencies determined from MC:
$(13.6\pm 0.2)\%$ for $K^-\pi^+$ and 
$(6.5\pm 0.2)\%$ for $K^-\pi^+\pi^+\pi^-$.

The branching fraction of $(D^*\to D\pi)\pi\pi$ events, calculated from the 
weighted average of the values obtained for the two modes, is:
$$
{\cal B}(B^-\to D^{*+}\pi^-\pi^-)=(1.25\pm0.08\pm0.22)\times10^{-3},
$$
where the first error is statistical and the second is systematic.
This  measurement is consistent with the world average value
$(2.1\pm0.6)\times 10^{-3}$~\cite{PDG}.
The systematic error is dominated by the uncertainties in the
track reconstruction efficiency $(16\,\%)$ (for a low momentum track 
from the  $D^*$ decay the efficiency
 uncertainty is $8\,\%$) and the 
PID efficiency $(5\,\%)$.
The background shape uncertainty is 
estimated in the same way
as for the $D\pi\pi$ analysis to be $5\%$.

\subsection{$B\to D^*\pi\pi$ coherent amplitude analysis}
In this final state we have a decaying vector $D^*$ particle.  
Assuming the width of the $D^*$ to be negligible, 
there are two additional degrees of freedom and, in addition to two $D^*\pi$ 
invariant masses,  two other variables
are needed to specify the final state.
The variables are chosen to be
the angle $\alpha$ between
the pions from the  $D^{**}$ and  $D^*$ decay in the 
$D^*$ rest frame, and the azimuthal angle $\gamma$ 
of the pion from the  $D^{*}$ 
relative to the $B\to D^*\pi\pi$ decay plane.

For further analysis, events satisfying  
the selection criteria described in the first section 
and having $\Delta E$ and $M_{\rm bc}$
within the 
$|\Delta E|<30~$MeV,  $|M_{\rm bc}-M_B|<6~{\rm MeV}/c^2$ signal range
are selected.
To understand the contribution and shape of the background, we use
events in the  $100~{\rm MeV}>|\Delta E|>30~{\rm MeV}$ sideband.

\begin{figure}[h]
\begin{center}
\begin{tabular}{cc}
\includegraphics[height=8 cm]{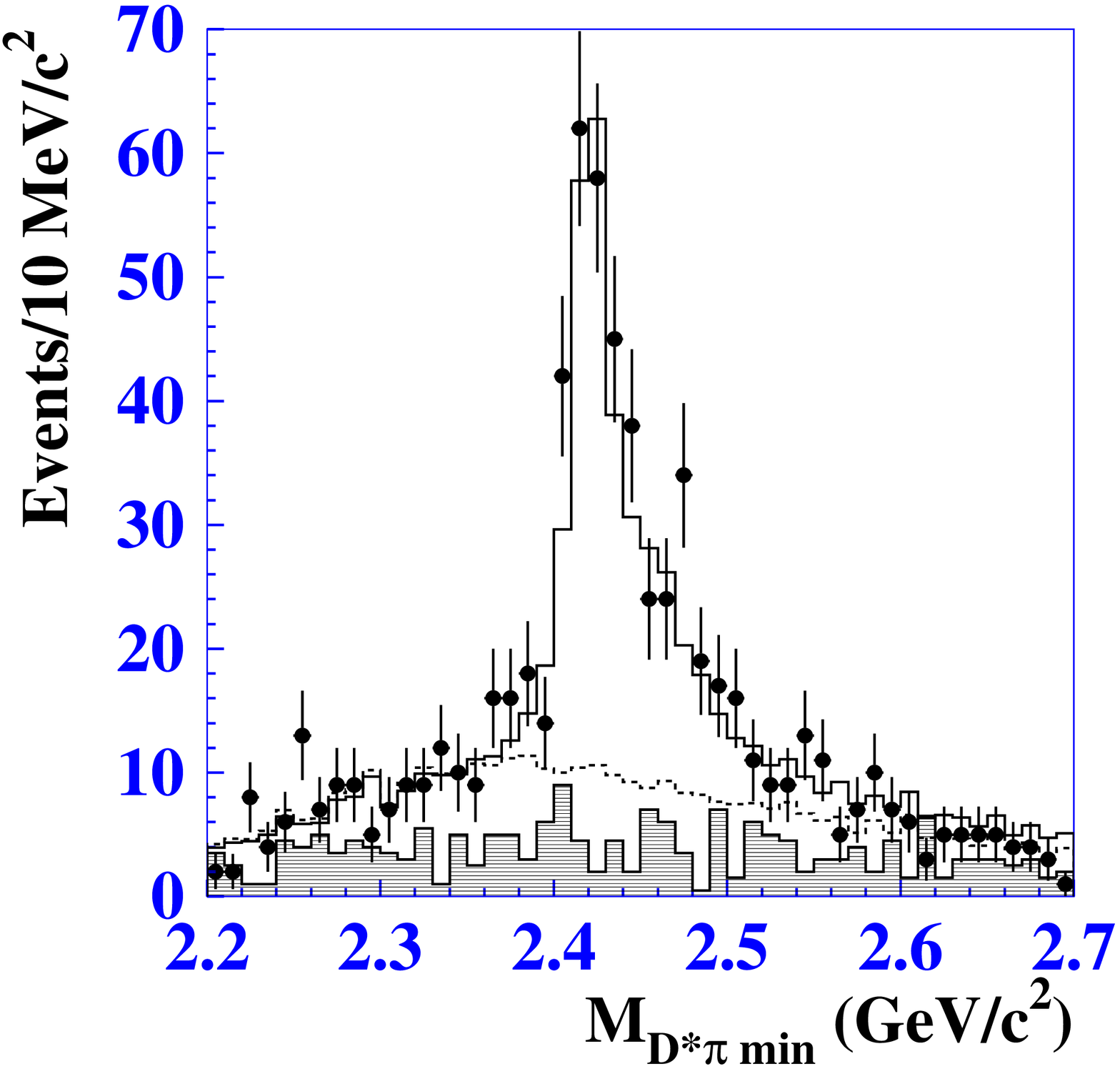}&
\includegraphics[height=8 cm]{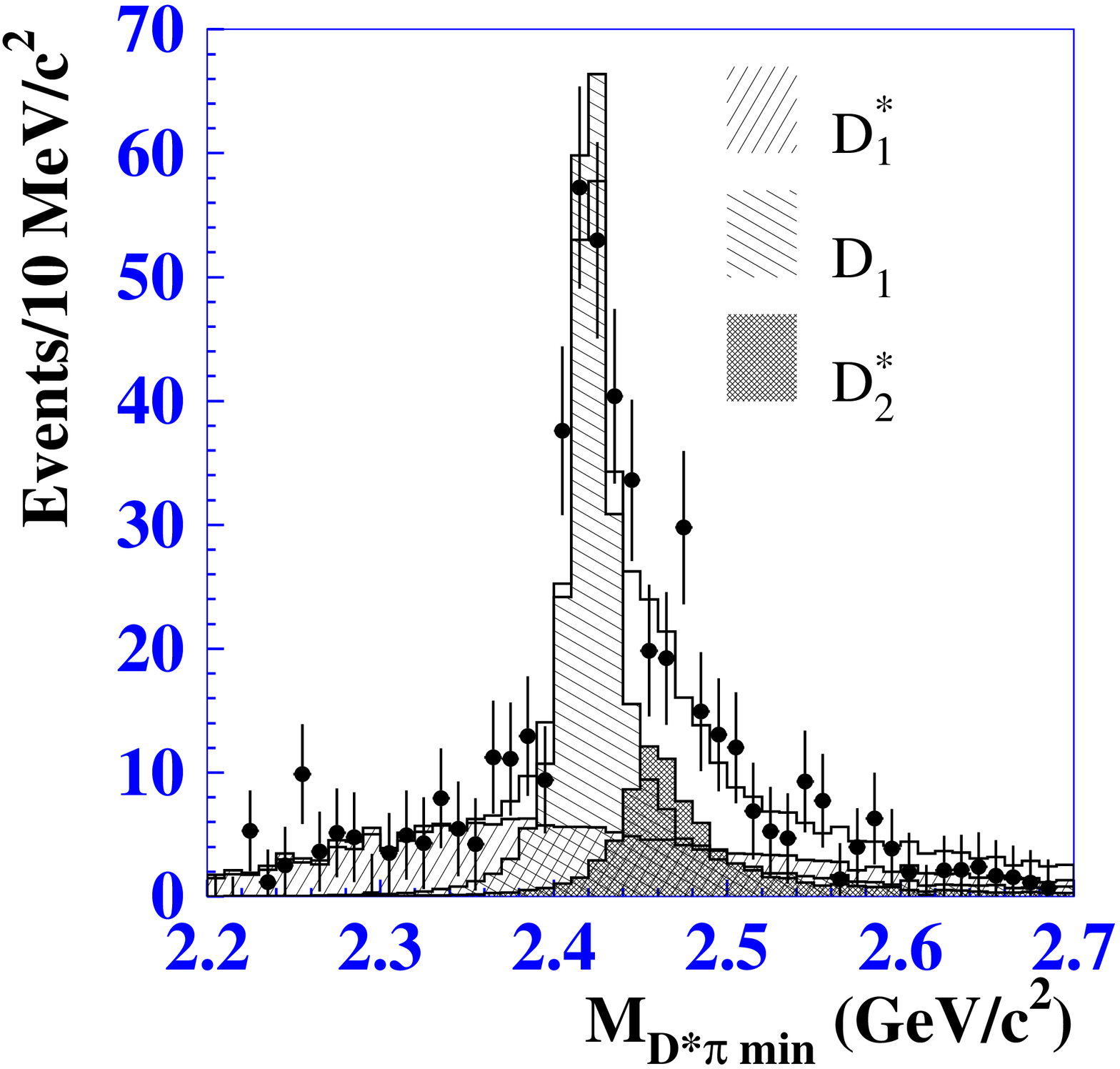}\\
\vspace*{-8 cm} & \\
{\bf\large \hspace*{-3cm} a)}&{\bf\large \hspace*{-3cm} b)}\\
\vspace*{6.5 cm} & \\
\end{tabular}
\caption{a) The minimal mass distribution of 
$D^*\pi$ events. 
The points with error bars are experimental data, 
the hatched histogram 
is the background distribution obtained from the sideband, 
the open histogram is MC simulation 
with the amplitudes and intermediate resonance parameters obtained from 
the fit. The dashed histogram shows the contribution of the broad
resonance.
b) The background-subtracted 
$D^*\pi$ mass distribution. The points  with error bars
correspond to 
the signal box events, hatched histograms show different
contributions, the open histogram  is a coherent sum of all contributions. }
\label{f:1_DS}
\end{center}
\end{figure}

The $D^* \pi~$ final state  can include contributions from the narrow 
$D_2^{*0}$ and $D^{0}_1$, and the broad $D'^{0}_1$
states.
The minimal $D^*\pi$ mass distributions for the signal and sideband events 
are shown in Fig.~\ref{f:1_DS}.
A narrow structure 
around $M_{D^*\pi}\sim 2.4~{\rm GeV}/c^2$ and a broader component that 
can be interpreted as the $D'_1$ are evident.

\begin{figure}[h]
\begin{center}
\begin{tabular}{cc}
\includegraphics[height=8 cm]{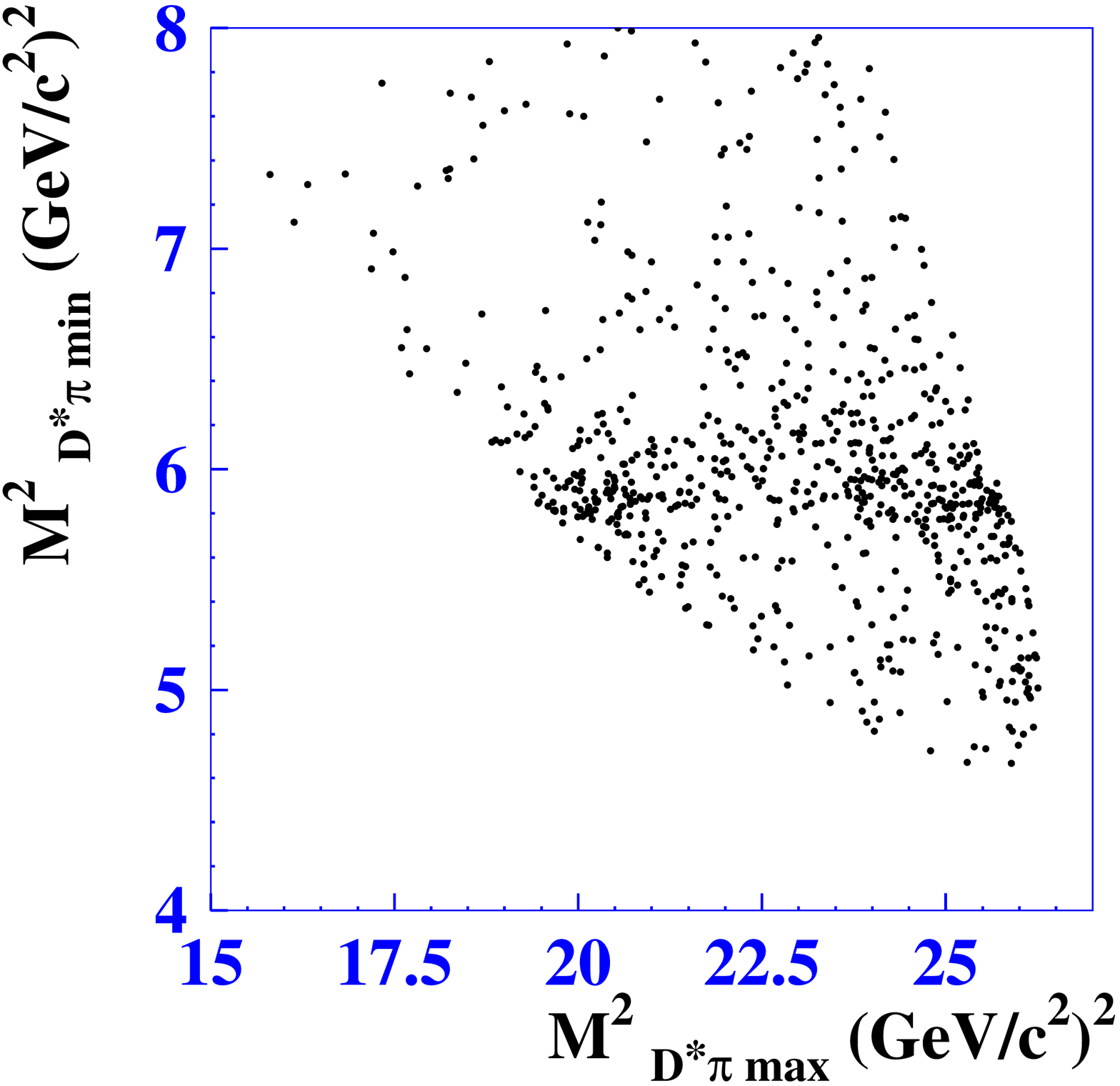}&
\includegraphics[height=8 cm]{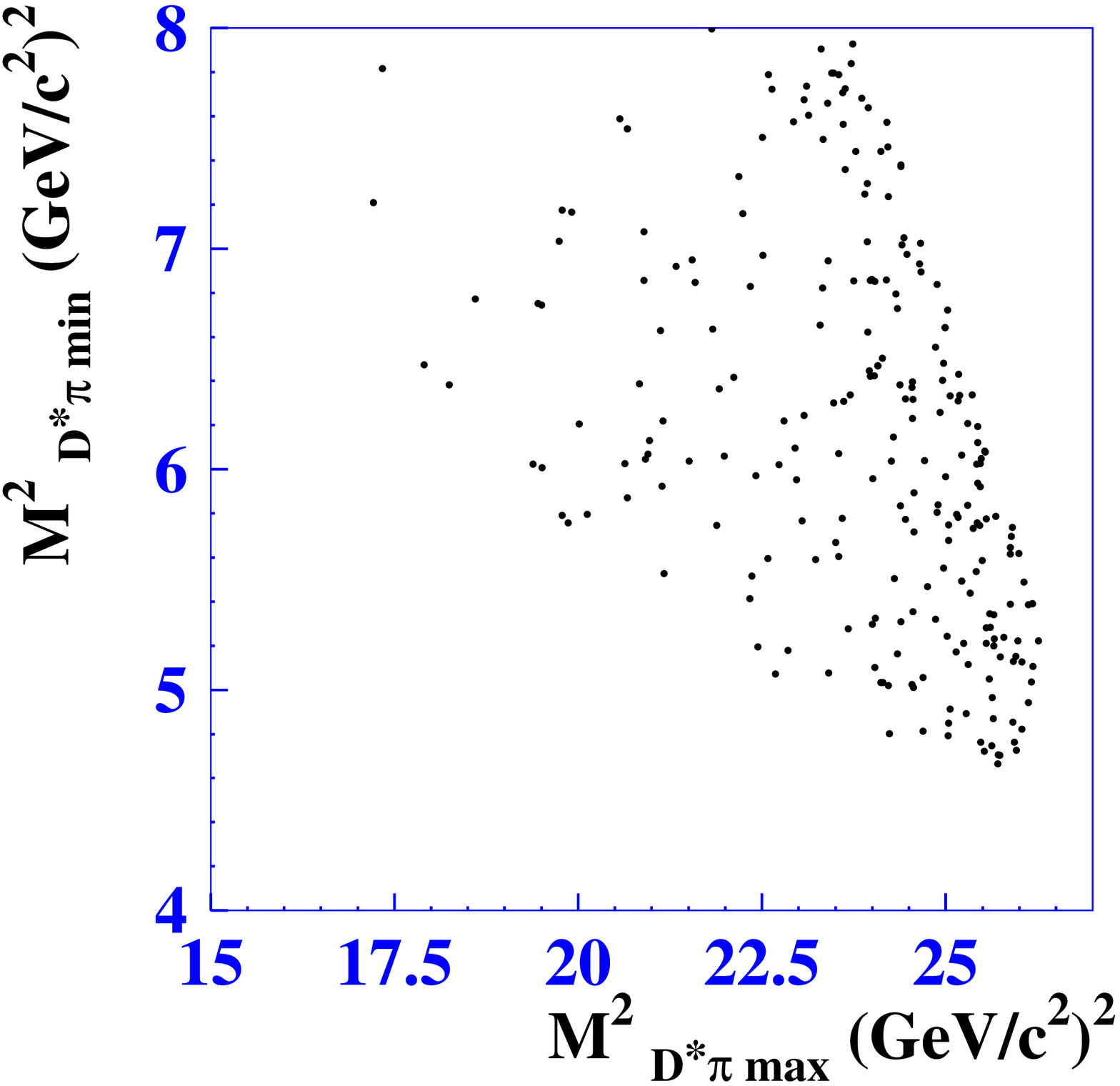}\\
\vspace*{-8.1 cm} & \\
{\hspace*{6.7cm}\bf\Large a)}&{\hspace*{6.7cm}\bf\Large b)}\\
\vspace*{6.6 cm} & \\
\end{tabular}
\caption{The Dalitz plot for a) signal and b) sideband events.}
\label{f:dpp_DS}
\end{center}
\end{figure}

The Dalitz plot distributions  
for the signal and sideband events
are shown 
in Fig.~\ref{f:dpp_DS}. 
In order to have the same boundary of the Dalitz plot distributions for 
events  from both signal and sideband regions as well as to decrease the 
smearing effect introduced by the detector resolution,
mass-constrained fits of $D\pi$ to $M_{D^{*+}}$ 
and $D^*\pi\pi$ to $M_B$ are performed. 

To extract the amplitudes and phases for different intermediate states, 
an  unbinned  likelihood  fit in 
the four-dimensional phase space  was performed. 
Assuming that the background distribution (${\cal
  B}(q^2_1,q^2_2,\alpha,\gamma)$)
in the signal region  has the 
same shape as in the $\Delta E$ sideband,
we obtain the 
${\cal B}(q^2_1,q^2_2,\alpha,\gamma)$ dependence from a fit of the sideband 
distribution to a smooth four-dimensional function.
The number of background events in the signal region 
is normalized according to the relative areas of the signal and the 
sideband regions. 
The signal is parameterized as a sum of the 
amplitudes of an intermediate tensor ($D_2^*$), and two axial vector 
mesons ($D'_1,~D_1$)
convoluted with 
the $q^2$ resolution function ${\cal{R}}(\Delta q^2)$ obtained from
MC simulation:
\begin{eqnarray}
S(q_1^2,q_2^2,\alpha,\gamma)&=&\big|
A^{(D_1)}(q_1^2,q_2^2,\alpha,\gamma)
+A^{(D'_1)}(q_1^2,q_2^2,\alpha,\gamma)+
A^{(D_2^*)}(q_1^2,q_1^2,\alpha,\gamma)\label{e:4.1}\\
&+&a_{D_v}e^{i\phi_{D_v}}A^{D_v}(q_1^2,q_2^2,\alpha,\gamma)+
a_{B^*_v}e^{i\phi_{B^*_v}}A^{B^*}(q_1^2,q_2^2,\alpha,\gamma)+
a_3e^{i\phi_3}\big|^2\otimes R(\Delta q^2).\nonumber
\end{eqnarray}

Each resonance is described by a relativistic Breit-Wigner with a width 
depending on $q^2$~(see Eq.\ref{e:brd}) and angular dependence
corresponding 
to the spins of the
intermediate and final state particles.
\begin{eqnarray}
T^{(1D)}(q_1,q_2,\alpha,\gamma)&=&a_{D_1}\frac{M_B^2\mathbf{p_{2}^2p_{1}p_{3}}}{\sqrt{q_1^2}}(\sin\theta\cos\gamma\sin\alpha+2\cos\theta\cos\alpha),\nonumber\\
T^{(1S)}(q_1,q_2,\alpha,\gamma)&=&a_{D'_1}e^{i\phi_{D'_1}}\frac{M_B\mathbf{p_{1}p_{3}}}{\sqrt{q_1^2}}(\sin\theta\cos\gamma\sin\alpha-\cos\theta\cos\alpha),\label{eqnms}\\
T^{(D_2^*)}(q_1,q_2,\alpha,\gamma)&=&a_{D^{*}_2}e^{i\phi_{D^{*0}_2}}\frac{M_B\mathbf{p_{2}^2p_{1}^2}p_{3}}{\sqrt{q_1^2}}\cos\theta\sin\theta\sin\alpha\sin\gamma,\nonumber
\end{eqnarray}
where $a_{D'_1},~a_{D_1},~a_{D^{*}_2},~\phi_{D'_1},~\phi_{D^{*}_2}$ 
are the relative amplitudes and phases for 
transitions via the
corresponding intermediate state.
The amplitudes of S and D waves in Eq.~(\ref{eqnms})
correspond to decay via   $1^+_{1/2}$ and $1^+_{3/2}$ intermediate 
states, respectively. 
Due to the finite c-quark mass, the observed $1^+$ states  can be a mixture 
of pure states.
Thus, the resulting amplitude will include a superposition of  
the amplitudes for the corresponding Breit-Wigner:
\begin{eqnarray}
T^{(D'_1)}(q_1,q_2,\alpha,\gamma)&=&T^{(1S)}(q_1,q_2,\alpha,\gamma)\cos\omega-e^{i\psi}
T^{(1D)}(q_1,q_2,\alpha,\gamma)\sin\omega\label{e:4.2}\\
T^{(D_1)}(q_1,q_2,\alpha,\gamma)&=&T^{(1S)}(q_1,q_2,\alpha,\gamma)\sin\omega+e^{-i\psi}
T^{(1D)}(q_1,q_2,\alpha,\gamma)\cos\omega .\nonumber
\nonumber
\end{eqnarray}
where  $\omega$ is the mixing angle and $\psi$ is a complex phase.

The $D^*\pi$ pair in the final state can be produced  via virtual $D^0_v$ or
$B^{*0}_v$ decaying to $D^{*+}\pi^-$.
Inclusion of a virtual  $D_v$ 
significantly improves the likelihood;
including in addition
$B^*_v$ and a constant term also improved the likelihood,  but 
the significance is not high (see Table~\ref{t:d1v1}).
\begin{table}[htb]
\begin{tabular}{|c|c|c|c|c|c|c|}
\hline
& $D_1,~D_2^*,~D'_1$
& $D_1,~D_2^*,~D'_1,D_v$&
& $D_1,~D_2^*,~D'_1,~D_v,$&
& $D_1,~D_2^*,~D'_1,~D_v,$
\\
&
& &
& $B^*_v$&
& $B^*_v,$  ph.sp$(a_3)$
\\
\hline
$Br_{D_1}(10^{-4})$    &   7.02 $\pm$  0.75  &   6.86 $\pm$  0.72    & &   6.78 $\pm$  0.69    & &   6.73 $\pm$  0.69     \\
$Br_{D_2^{*}}(10^{-4})$&   1.89 $\pm$  0.28  &   2.00 $\pm$  0.28    & &   1.83 $\pm$  0.26    & &   1.82 $\pm$  0.27     \\
$\phi_{D_2^{*}}$       &  -0.53 $\pm$  0.15  &  -0.56 $\pm$  0.14    & &  -0.57 $\pm$  0.14    & &  -0.56 $\pm$  0.14     \\
$Br_{D'_1}(10^{-4})$&   5.01 $\pm$  0.40  &   4.99 $\pm$  0.39    & &   4.96 $\pm$  0.38    & &   4.84 $\pm$  0.38     \\
$\phi_{D'_1}$       &   1.86 $\pm$  0.18  &   1.65 $\pm$  0.23    & &   1.68 $\pm$  0.20    & &   1.70 $\pm$  0.20     \\
$Br_{D_v}(10^{-4})$    &   --                &   0.52 $\pm$  0.19    & &   0.57 $\pm$  0.19    & &   0.57 $\pm$  0.19     \\
$\phi_{D_v}$           &  --                 &  -2.68 $\pm$  0.26    & &  -2.43 $\pm$  0.24    & &  -2.43 $\pm$  0.25     \\
$Br_{B_v^{*}}(10^{-4})$&   --                &   --                  & &   0.21 $\pm$  0.10    & &   0.21 $\pm$  0.11     \\
$\phi_{B_v^{*}}$       &   --                &   --                  & &   1.19 $\pm$  0.44    & &   1.23 $\pm$  0.43     \\
$M_{D_1^0   }(MeV/c^2)$& 2421.4 $\pm$   1.6  & 2421.2 $\pm$   1.5    & & 2421.4 $\pm$   1.5    & & 2421.3 $\pm$   1.5     \\
$\Gamma_{D_1^0}(MeV)$  &   26.7 $\pm$   3.1  &   25.2 $\pm$   2.9    & &   23.7 $\pm$   2.7    & &   23.5 $\pm$   2.8     \\
$M_{D'^{0}_1}(MeV/c^2)$&  2442  $\pm$   29   &  2433  $\pm$   29     &
&  2427  $\pm$   26     & &  2425  $\pm$   26      \\
$\Gamma_{D'^{0}_1}(MeV)$&   454  $\pm$  100   &   417  $\pm$  105
& &   384  $^{+107}_{-75}$ & &   374  $\pm$   87      \\
$\omega$               &  -0.08 $\pm$  0.03  &  -0.09 $\pm$  0.03    & &  -0.10 $\pm$  0.03    & &  -0.10 $\pm$  0.03     \\
$\psi$                 &   0.00 $\pm$  0.22  &   0.05 $\pm$  0.21    & &   0.05 $\pm$  0.20    & &   0.06 $\pm$  0.20     \\
$a_3\times 10^4 $          &  -- &   --   &&    --  && $0.51   \pm
0.77$        \\
$\phi_3 $          &  -- &   --   &&    --  && $-0.08  \pm 0.83$        \\
$N_{sig2}$             &   277  $\pm$   21   &   274  $\pm$   20     &
&   279  $\pm$   20     & &   278  $\pm$   20      \\
$N_{sig4}$             &   275  $\pm$   20   &   276  $\pm$   20     &
&   281  $\pm$   20     & &   281 $\pm$   20      \\
\hline
$-2\ln\cal{L/L_0}$     &      25           &      7
&&       0           & &      -2         \\
\hline                                                                         \end{tabular}
\caption{Fit results for different models.
The model that is used to obtain these results includes amplitudes for
$D^*_2,~D_1,~D'_1,D_v,~B^*_v$ intermediate resonances. Adding a constant
 term does not improve the likelihood significantly.
}
\label{t:d1v1}
\end{table}
A fit without the inclusion of a broad 
resonance gives a considerably worse likelihood (see
Table~\ref{t:d1v}). 
We also tried
to fit the data by including a  broad resonance 
with other quantum numbers such as $0^-,1^-,~2^+$. In these cases
the likelihood is also significantly worse, as shown in Table~\ref{t:d1v}.
We conclude that we have observed the broad $1^+$~$D'_1$ state with
a statistical significance of more than 10$~\sigma$.
The model and systematic errors
 are estimated in the same way as for the $D\pi\pi$ case.

The $D^{*}_2$  mass and width are 
fixed to the values obtained from the $D\pi\pi$ analysis. The axial
vector $D^{**}$
masses and widths as well as the branching fractions and 
phases of amplitudes 
$a_{D_1},~a_{D'_1},~a_{D^{*}_2},~\phi_{D'_1},~\phi_{D^{*}_2}$ are 
treated as free parameters of the 
fit as are the mixing angle $\omega$ and the mixing phase $\psi$. 

\begin{figure}[phtb]
\begin{tabular}{ccc}
\includegraphics[height=4.5 cm]{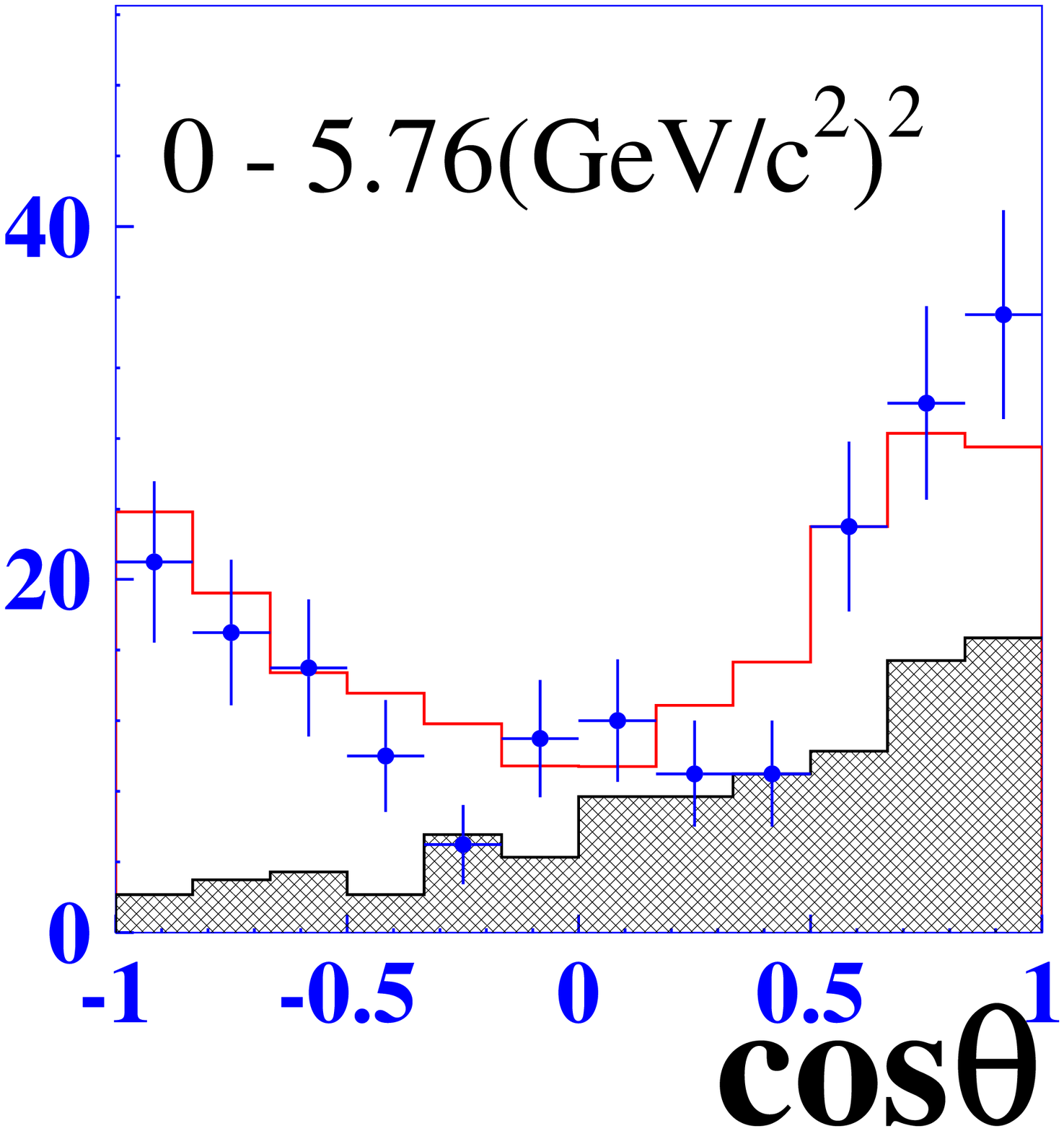}&
\includegraphics[height=4.5 cm]{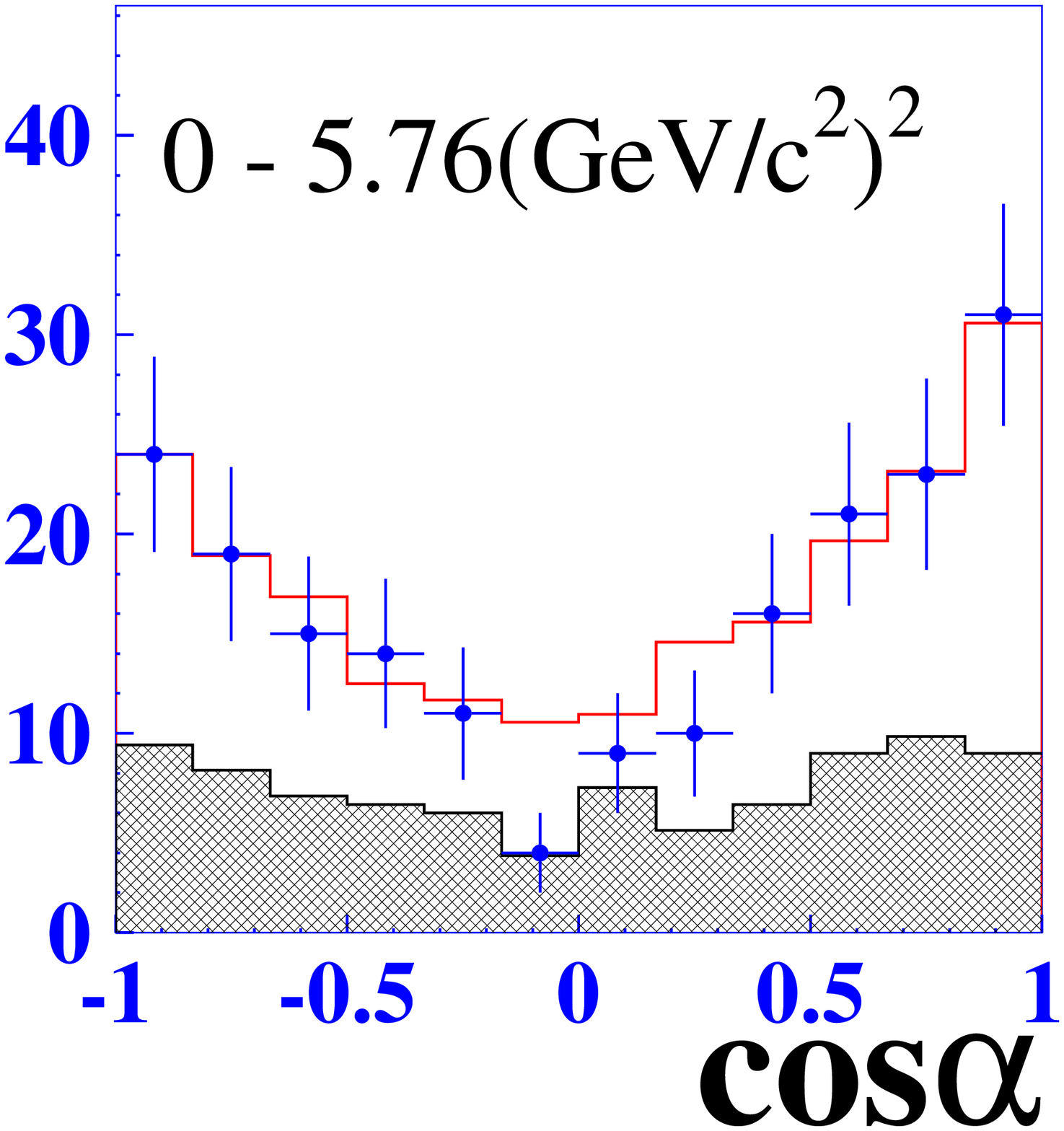}&
\includegraphics[height=4.5 cm]{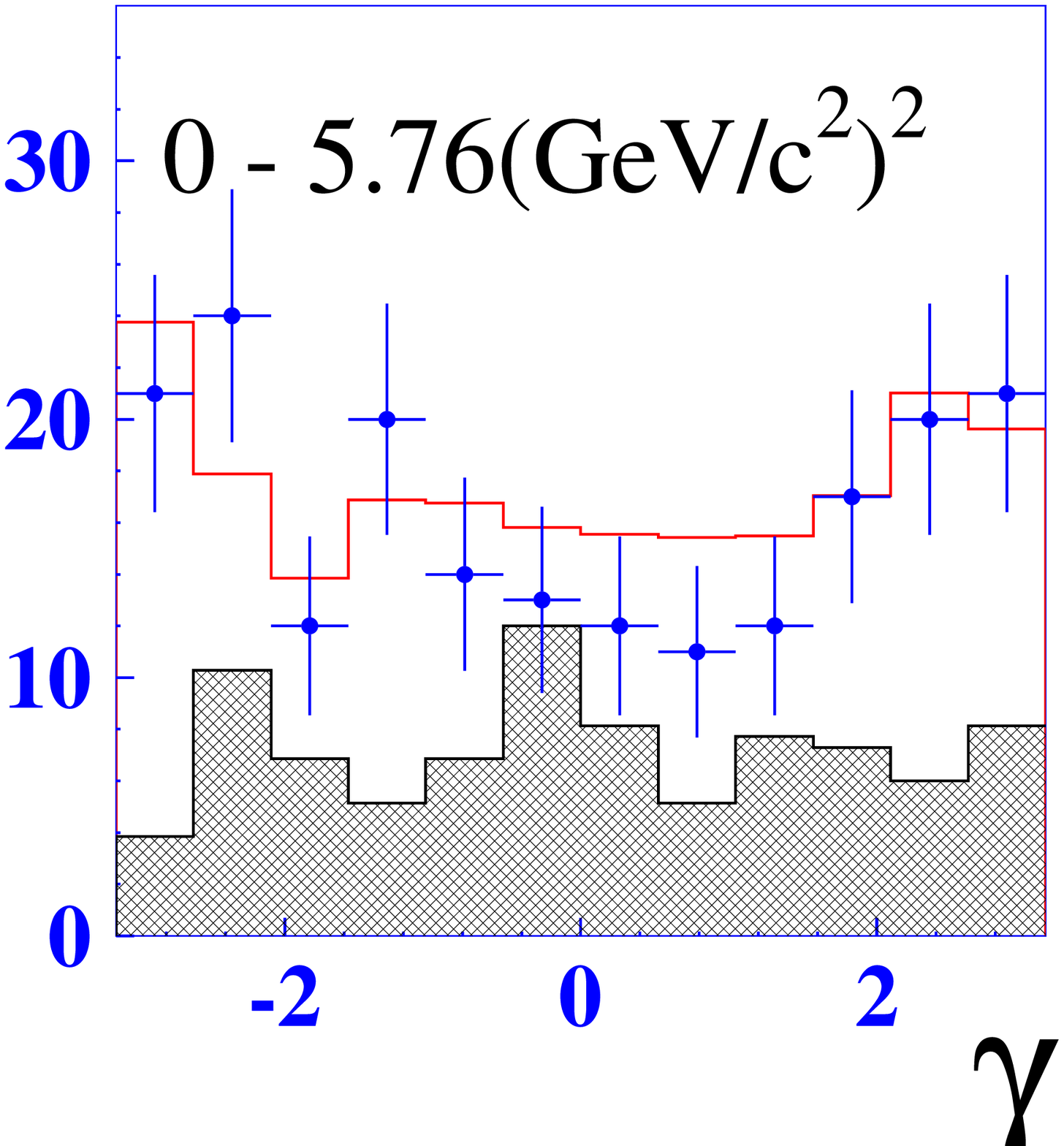}\\
\includegraphics[height=4.5 cm]{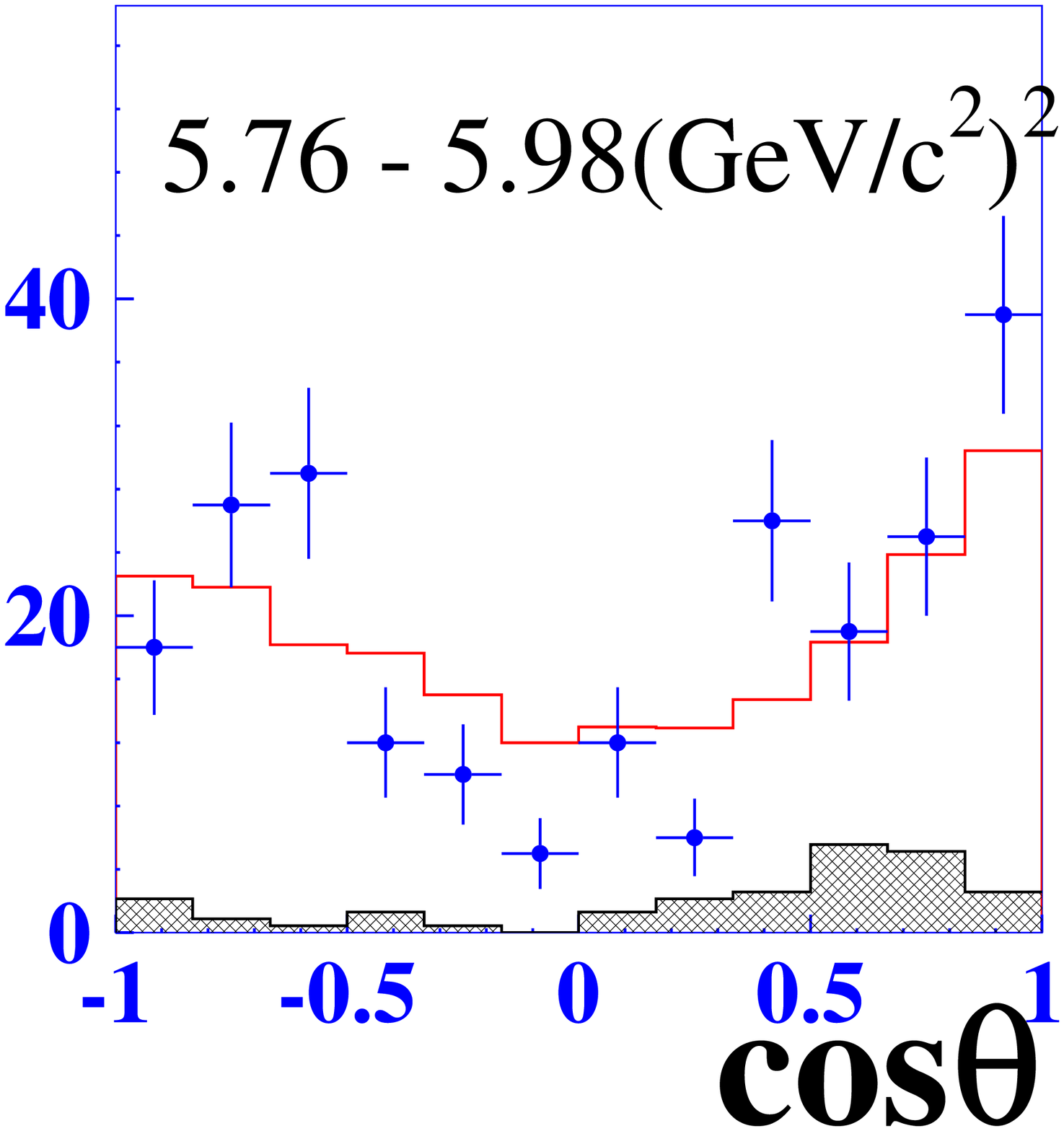}&
\includegraphics[height=4.5 cm]{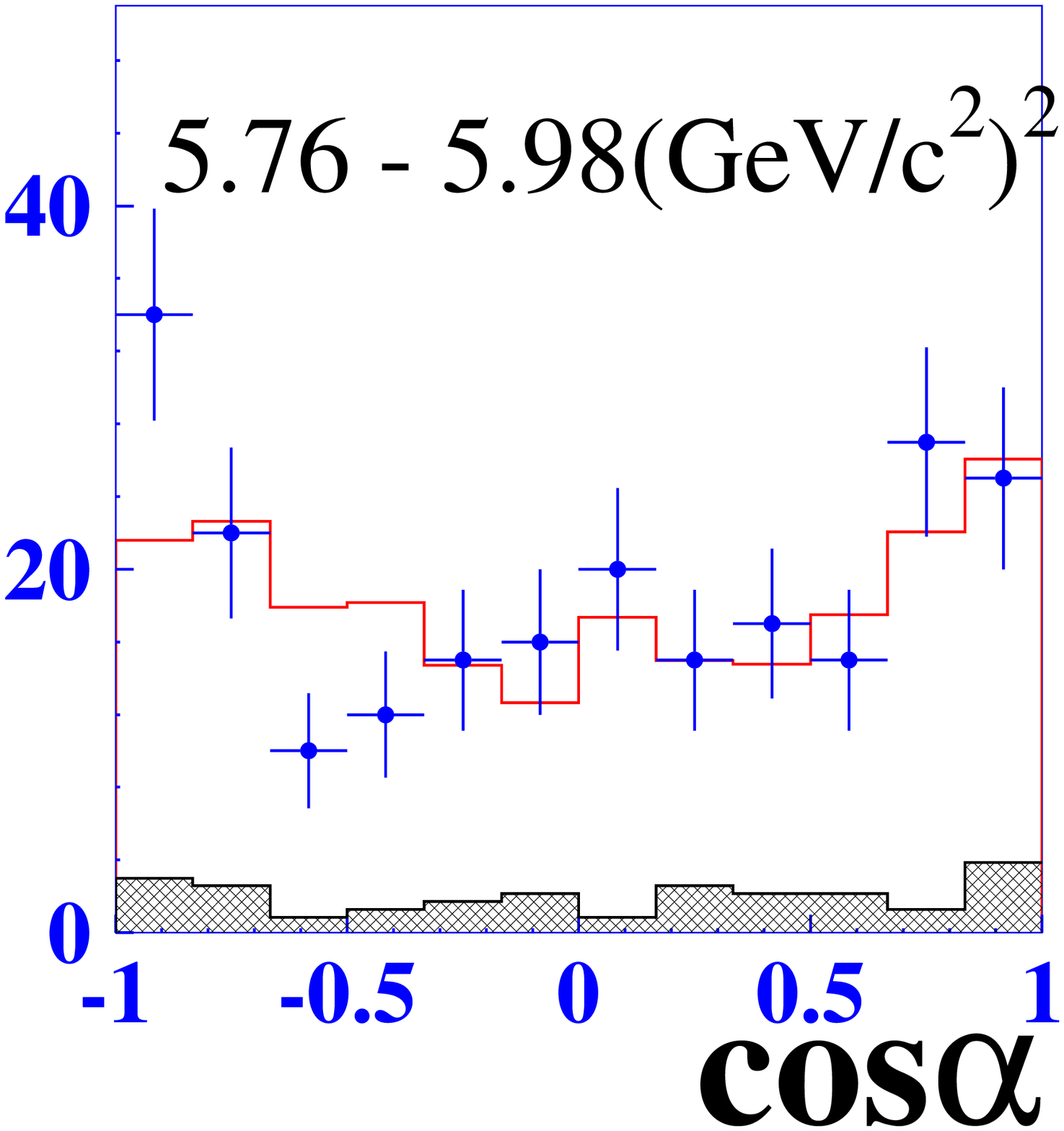}&
\includegraphics[height=4.5 cm]{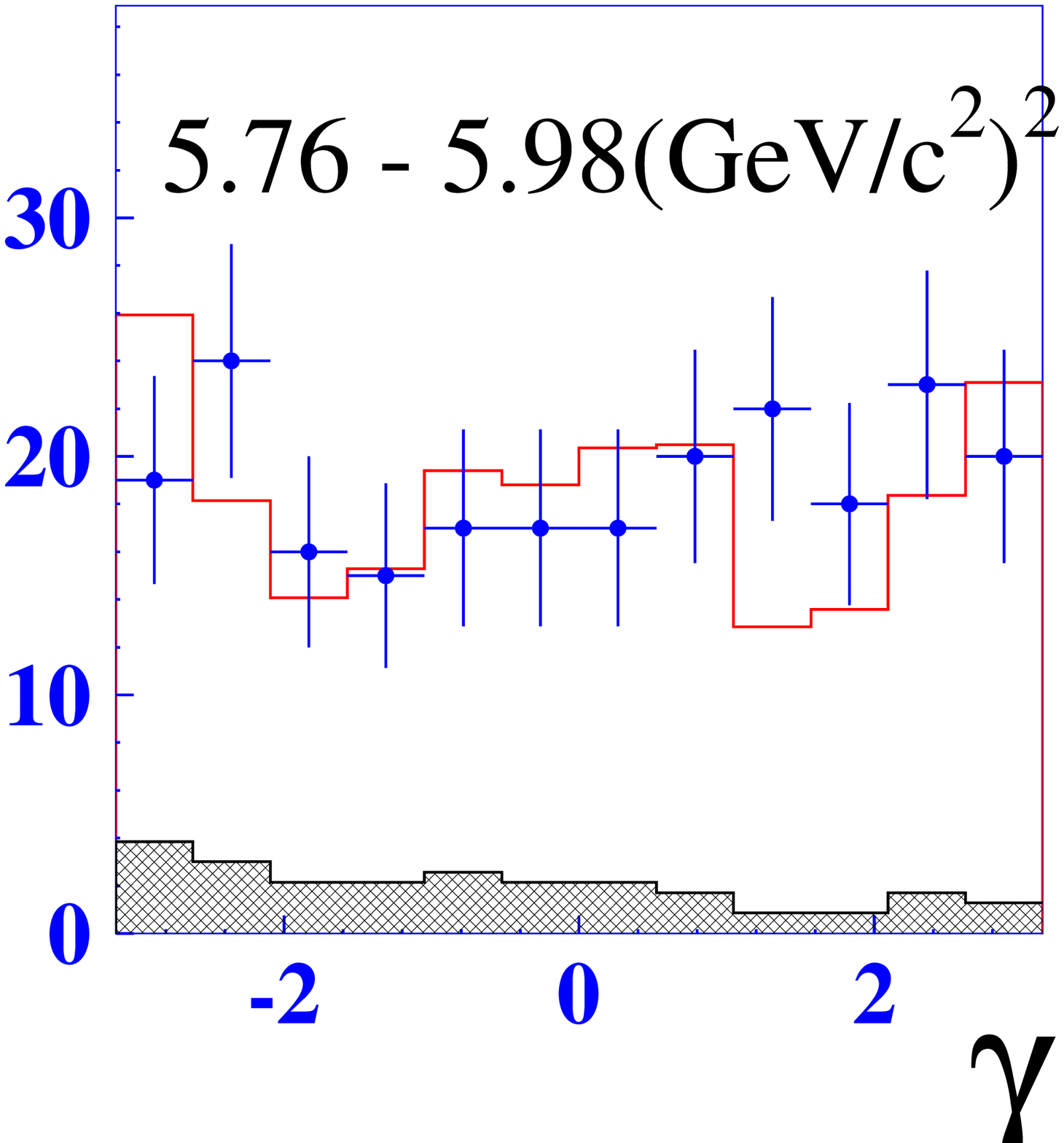}\\
\includegraphics[height=4.5 cm]{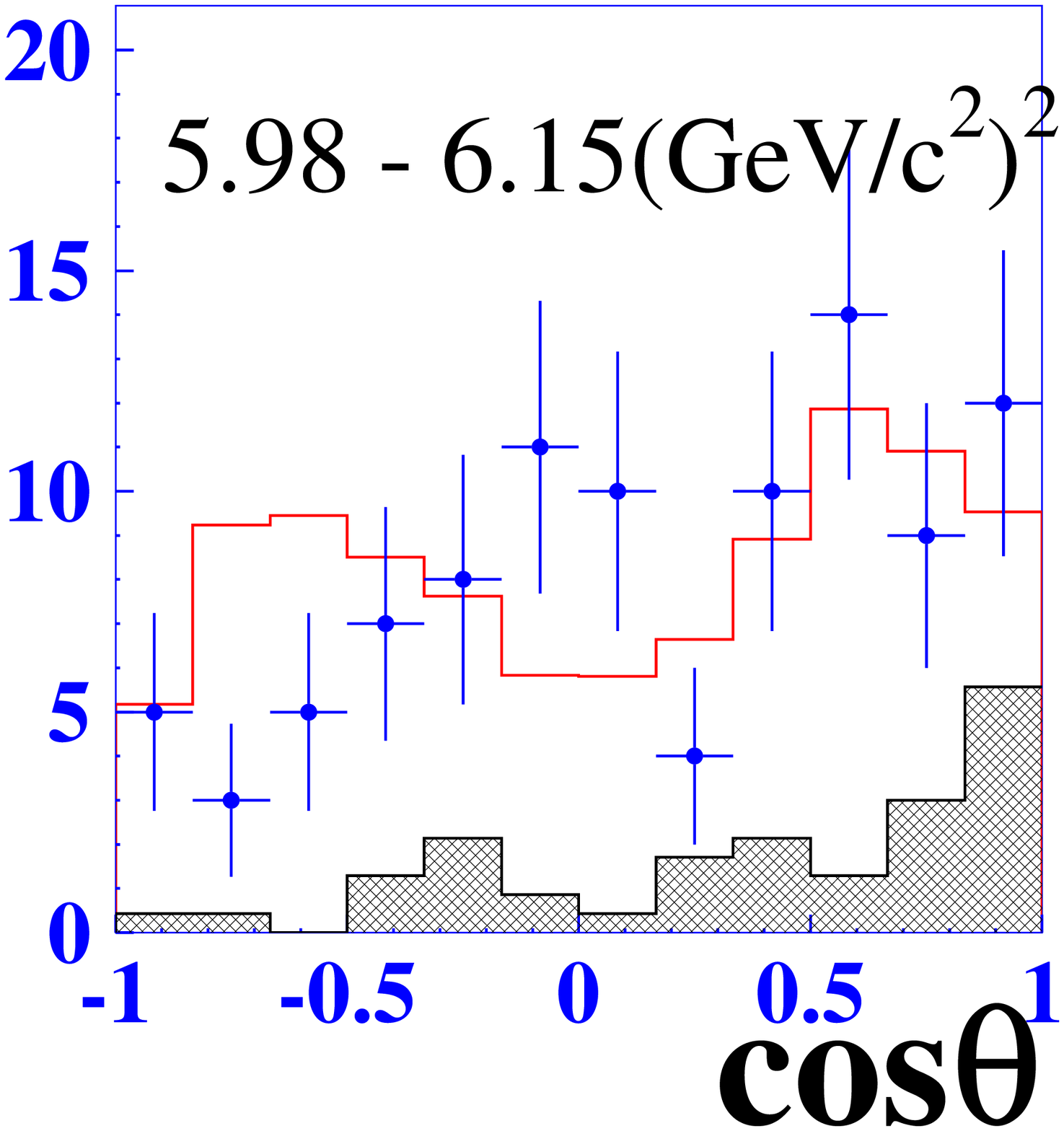}&
\includegraphics[height=4.5 cm]{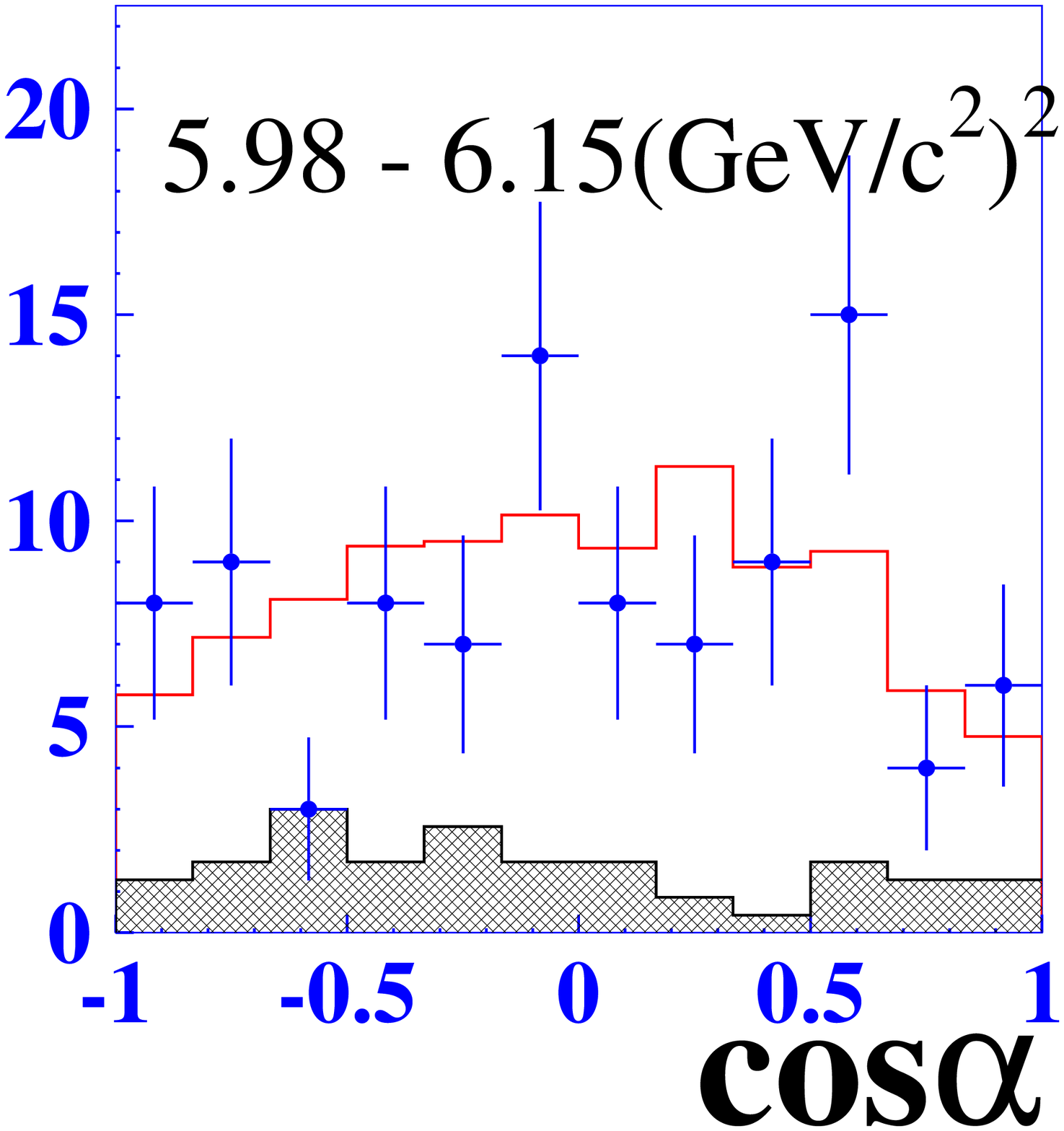}&
\includegraphics[height=4.5 cm]{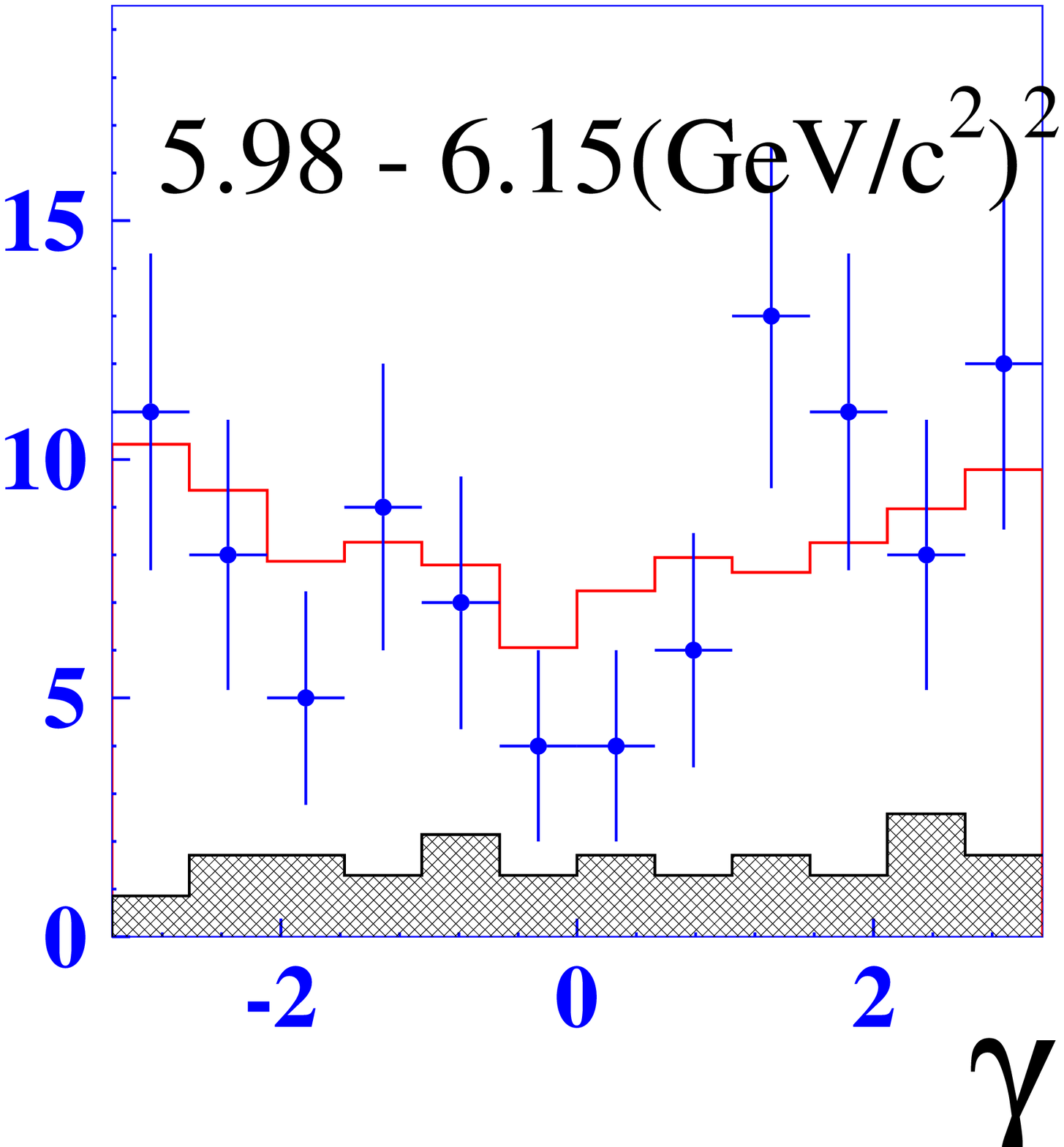}\\
\includegraphics[height=4.5 cm]{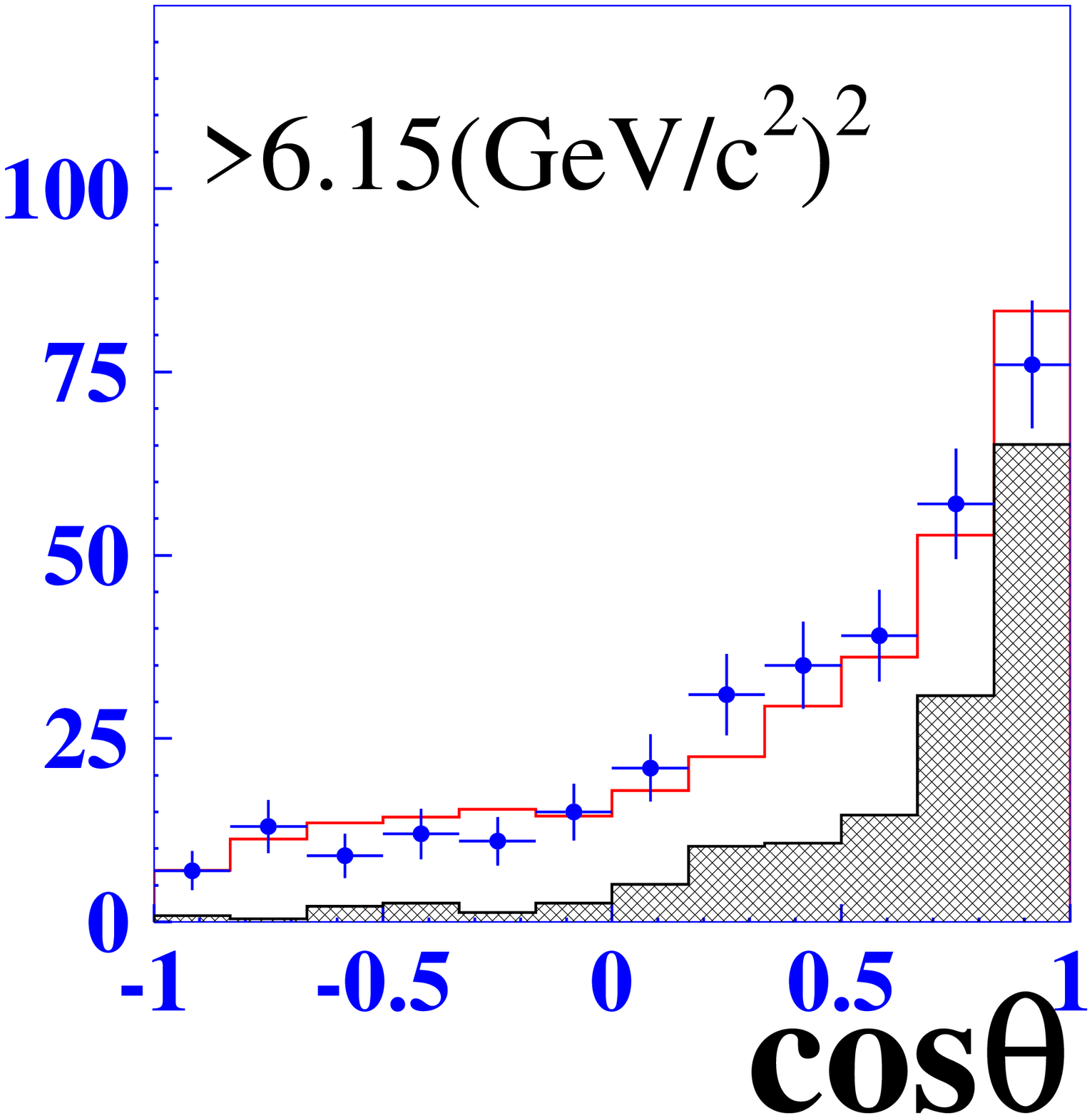}&
\includegraphics[height=4.5 cm]{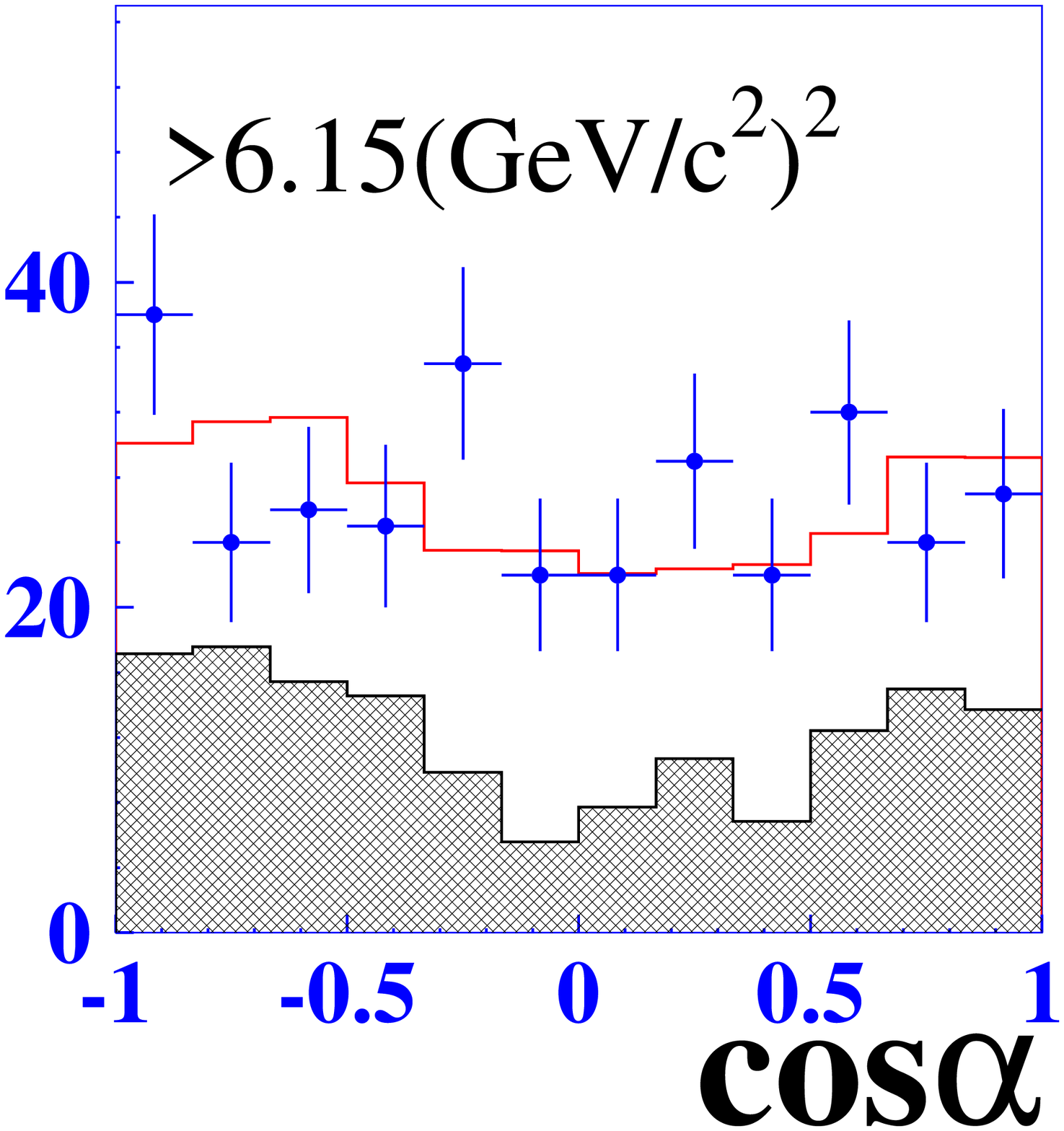}&
\includegraphics[height=4.5 cm]{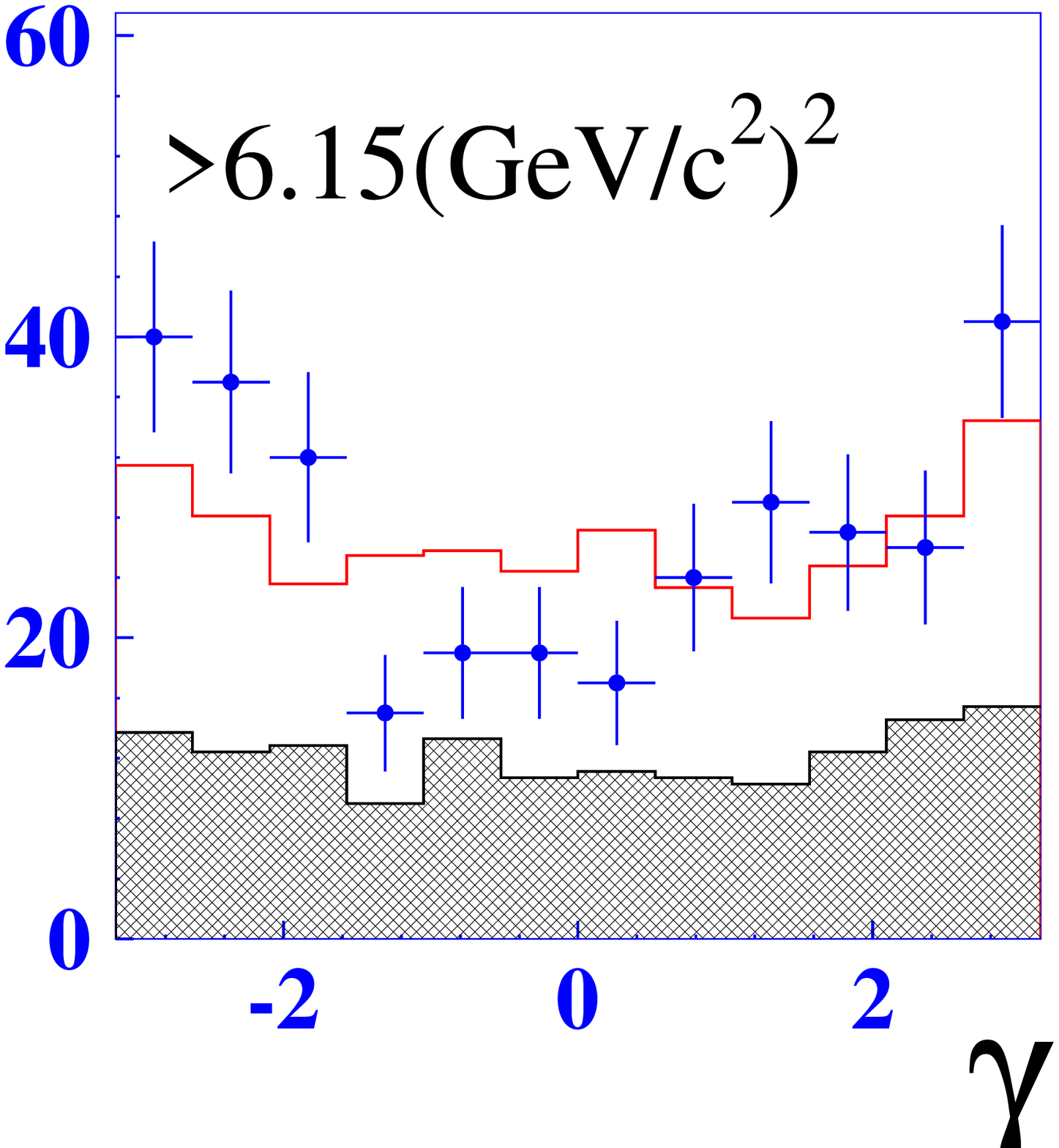}\\
\vspace*{-18.2 cm} & &\\
{\hspace*{-1.cm}\bf\Large a)}&{\hspace*{-1.cm}\bf\Large b)}&{\hspace*{-1.cm}\bf\Large c)}\\
\vspace*{3.6 cm} & & \\
{\hspace*{-1.cm}\bf\Large d)}&{\hspace*{-1.cm}\bf\Large
  e)}&{\hspace*{-1.cm}\bf\Large f)}\\
\vspace*{3.6 cm} & & \\
{\hspace*{-1.cm}\bf\Large f)}&{\hspace*{-1.cm}\bf\Large
  g)}&{\hspace*{-1.cm}\bf\Large h)}\\
\vspace*{3.6 cm} & & \\
{\hspace*{-1.cm}\bf\Large i)}&{\hspace*{-1.cm}\bf\Large
  j)}&{\hspace*{-1.cm}\bf\Large k)}\\

\vspace*{3.6 cm} & & \\
\end{tabular}
\caption{Distributions of the
helicity angle of $D^{**}$($\cos\theta$), 
helicity angle of $D^{*}$($\cos\alpha$) and 
azimuthal angle $\gamma$ in four
 different $q^2$ regions.
The points are the experimental data, 
the histogram is MC simulation with fitted parameters, 
the hatched histogram shows the background contribution (from the $\Delta E$ sideband).}
\label{f:dps_h_d2}
\end{figure}

Since
there is no good way to graphically present the data and the
model in four dimensions, we show the projections of the distributions for 
various variables.
Figure~\ref{f:1_DS} shows the $M_{D^*\pi~min}$ distribution together
with MC results that were generated 
according to the model containing $D_1,~D'_1,~D_2^*$ and virtual
$D_v,~B^*_v$
 intermediate resonances 
with parameters obtained from the fit.
Figure~\ref{f:dps_h_d2} shows  
a comparison of the data and the MC simulation for 
 $D^{**}$ and $D^{*}$ helicities as well as the angle $\gamma$ for  
$q^2$ ranges corresponding to  the two narrow resonances
$D_1$ ($q^2=(5.76\sim5.98)\rm\,(GeV/c^2)^2$), $D_2^*$ ($q^2=(5.98\sim6.15)
\rm\,(GeV/c^2)^2$) and the regions populated mainly by the broad
$D'_1$ state  below
 ($q^2<5.76~\rm\,(GeV/c^2)^2$) and above 
($q^2>6.15~\rm\,(GeV/c^2)^2$) the narrow resonances.
All distributions indicate good agreement between the data and the fit
result.
We cannot characterize the 
quality of the fit by the standard $\chi^2$ test since for a binned
distribution with four degrees of freedom and a limited data sample
any reasonable binning will result in only a few  events per bin. 
Therefore to estimate the quality of the fit we determine $\chi^2$
values for different projections of the distributions in
Figs.~\ref{f:1_DS} and \ref{f:dps_h_d2}. The obtained $\chi^2$ values
correspond to confidence levels in the  $5-90$\,\% range.

For  the $D_1$ meson we obtain the following parameters:
$$
M_{D^{0}_1}=(2421.4\pm1.5\pm0.4\pm0.8)\, {\rm MeV}/c^2,~\Gamma_{D^{0}_1}=(23.7\pm2.7\pm0.2\pm4.0)\, \rm{MeV}.
$$
These parameters  are in good agreement with the world average values:
$
M_{D^{0}_1}=(2422.2\pm1.8)\, {\rm MeV}/c^2,~\Gamma_{D^{0}_1}=(18.9^{+4.6}_{-3.5})\, \rm{MeV}
$~\cite{PDG}.

The broad $D'^{0}_1$ resonance parameters are:
$$
M_{D'^{0}_1}=(2427\pm26\pm20\pm15)\, {\rm MeV}/c^2,~\Gamma_{D'^{0}_1}=(384^{+107}_{-75}\pm24\pm70)\, {\rm MeV}.
$$
Observation of a similar state was reported by CLEO but was 
not published;
our measurement is consistent with CLEO's preliminary results:
$M_{D'^{0}_1}=(2461^{+48}_{-42})\,{\rm MeV}/c^2,
~\Gamma_{D'^{0}_1}=(290^{+110}_{-90})\,{\rm MeV}$~\cite{CLEO996}.

The results for the products of the branching fractions of the  
$B$ and $D^{**}$ mesons are:
$$
{\cal B}(B^-\to D_1\pi^-)\times {\cal B}(D_1\to D^{*+}\pi^-)=(6.8\pm0.7\pm1.3\pm0.3)\times10^{-4},
$$
$$
{\cal B}(B^-\to D^{*0}_2\pi^-)\times {\cal B}(D_2^{*0}\to D^{*+}\pi^-)=(1.8\pm0.3\pm0.3\pm0.2)\times10^{-4},
$$
$$
{\cal B}(B^-\to D'^{0}_1\pi^-)\times {\cal B}(D'^{0}_1\to D^{*+}\pi^-)=(5.0\pm0.4\pm1.0\pm0.4)\times10^{-4},
$$
the relative phases of the $D^*_2$ and $D'_1$ amplitudes are:
$$
\phi_{D^{*0}_2}=-0.57\pm0.14\pm0.06\pm0.13 ;~\phi_{D'_1}=1.68\pm 0.20\pm0.07\pm0.16, 
$$
and the mixing angle of two axial states and the complex phase are:
$$
\omega=-0.10\pm 0.03\pm0.02\pm0.02,
$$
$$
\psi=0.05\pm0.20\pm0.04\pm0.06.
$$

To  understand the uncertainties in the background shape and
the efficiency of the cuts, additional studies are performed.
The background shapes obtained separately 
for the upper and lower $\Delta E$ sidebands
are used in the likelihood optimization. 
We also apply  more restrictive  cuts on $\Delta E$ and $M_{\rm bc}$ 
that improve
the signal-to-background ratio by about a factor of two and repeat the fit.
The maximum difference between the values obtained with 
different cuts and different background shapes is included in the
systematic error.
The branching fraction errors also include an 18\% systematic 
uncertainty in the detection efficiency.
The model uncertainties are estimated by comparing fit results for the case 
of different models (Table~\ref{t:d1v1}) and for values of $r$
in the range from 0 to 5 (GeV/c)$^{-1}$, where $r$ is the hadron scale
parameter in  the transition form factors 
of Eqs.~(\ref{eqfo}) and (\ref{eqf1}).
\begin{table}
\begin{center}
\begin{tabular}{|c|c|}
\hline
Model & $-2\ln\cal{L/L}_0$ \\
\hline
${D^{\star}_2,D'_1,D_1,D_v,B^{\star}_v}$ & 0\\
\hline                           
${D^{\star}_2,D_1,D_v,B^{\star}_v}$,  ph.sp$(a_3)$ & 170 \\
${D^{\star}_2,D_1,D_v,B^{\star}_v,0^-}$ & 107 \\
${D^{\star}_2,D_1,D_v,B^{\star}_v,1^-}$ &156\\
${D^{\star}_2,D_1,D_v,B^{\star}_v,2^+}$ & 166\\
\hline
\end{tabular}
\caption{Comparison of the models with and without a broad $1^+$ resonance.
The $D_2^*$ and  $D_1$ amplitudes are always included. }
\label{t:d1v}
\end{center}
\end{table}

\section{Discussion}

From the measured products of branching fractions of  
${\cal B}(B^-\to D^{*0}_2\pi^-){\cal B}(D_2^{*0}\to D^{*+}\pi^-)$ 
and ${\cal B}(B^-\to D^{*0}_2\pi^-){\cal B}(D_2^{*0}\to D^{+}\pi^-)$ we obtain the 
ratio of the $D^{*0}_2$ branching fractions:
$$
H=\frac{{\cal B}(D_2^{*0}\to D^{+}\pi^-)}{{\cal B}(D_2^{*0}\to D^{*+}\pi^-)}=1.9\pm0.5,
$$
which is consistent with the world average $H=2.3\pm 0.6$.
Theoretical models~\cite{rosner,godfrey,falk} predict $H$ to be 
in the range from 1.5 to 3.  
 If the $D^{*}_2$ decay
is saturated by  $D\pi$, $D^*\pi$ transitions, and $D_1$ decay by
$D^*\pi$,
then  the ratio 
$R~$ in Eq.~(\ref{e:Neu1}) can be expressed as  the following
combination of branching fractions:
$$
R=\frac{{\cal B}(B^-\to D^{*0}_2\pi^-)({\cal B}(D_2^{*0}\to D^{*+}\pi^-)+{\cal B}(D_2^{*0}\to D^{+}\pi^-))}{{\cal B}(B^-\to D^0_1\pi^-){\cal B}(D_1^{0}\to D^{*+}\pi^-)}=0.77\pm 0.15.
$$
The obtained value is lower than that of the CLEO measurement
(although the measurements are consistent within errors) but is 
still a factor of two larger than the  factorization result~\cite{neubert}.
From our measurement
it is impossible to determine  whether the 
non-factorized part for  tensor and axial mesons is large, or whether 
higher order corrections to the leading  factorized terms
should be taken into account.
According to Ref.~\cite{leg}, the observed value of $R$ corresponds to
a value of the sub-leading Isgur-Wise 
function $\hat{\tau}_1=0.40^{+0.10}_{-0.15}$~GeV.

For semileptonic decays, where there is no non-factorized contribution, 
the corresponding ratio  is $0.5\pm0.6  $~\cite{ALEPH}, 
which,  within experimental 
errors, is consistent with both our measurement and 
the model prediction. More accurate measurements of
semileptonic modes containing $D^{**}$ mesons may help  resolve this 
problem.

Our measurements show that the narrow resonances  
comprise  $(36\pm6)\,\%$ of 
$D\pi\pi$ decays and $(63\pm6)\,\%$ of $D^*\pi\pi$ decays.
This  result is inconsistent with the QCD sum rule~\cite{QCDSR} that 
predicts 
the dominance of the narrow states 
in $B\to D^{(*)}\pi\pi$ decays. It is also possible  that
in $B^{-}\to D^{**0}\pi^-$ decays the color suppressed amplitude is
comparable to the tree amplitude, so that other transition
form factors play a role.  
The ratio of the production rates for narrow and broad $D\pi$ states in 
semileptonic $B\to D^{(*)}\pi l\nu$  decays
measured at LEP~\cite{ALEPH} also indicates an excess of the  broad states.
More accurate measurements of both semileptonic decays and other charged 
states of the $D^{(*)}\pi\pi$ 
system may resolve this discrepancy.

\section{Conclusion}

 We have performed a study of charged $B\to D^+\pi^-\pi^-$ and 
$B\to D^{*+}\pi^-\pi^-$ decays.
The total branching fractions  have been measured
to be
${\cal B}(B^-\to D^+\pi^-\pi^-)=(1.02\pm0.04\pm0.15)\times10^{-3}$ and 
${\cal B}(B^-\to D^{*+}\pi^-\pi^-)=(1.25\pm0.08\pm0.22)\times10^{-3}$. 
For the former decay this is the first measurement.

A study of the dynamics of these three-body decays is reported.
The $D^+\pi^-\pi^-$ final state is well described by the 
production of $D^*_2\pi^-$ and $D^*_0\pi^-$ followed by $D^{**}\to D\pi$.
From a Dalitz plot analysis we obtain the mass, width and  
product of the branching fractions for the $D^{*0}_2$: 
$$
M_{D^{*0}_2}=(2461.6\pm2.1\pm0.5\pm3.3)\, {\rm MeV}/c^2,~\Gamma_{D^{*0}_2}=(45.6\pm4.4\pm6.5\pm1.6)\,{\rm MeV},
$$
$$
{\cal B}(B^-\to D^{*0}_2\pi^-)\times {\cal B}(D_2^{*0}\to D^{+}\pi^-)=(3.4\pm0.3\pm0.6\pm0.4)\times10^{-4}.
$$
In this mode we also observe production of a broad scalar $D^*_0$ meson with 
 mass and width:
$$
M_{D^{*0}_0}=(2308\pm17\pm15\pm28)\, {\rm MeV}/c^2, \Gamma_{D^{*0}_0}=(276\pm21\pm18\pm60)\,{\rm MeV}.
$$
The product of the   branching fractions  for the $D_0^*$ state is
$$
{\cal B}(B^-\to D^{*0}_0\pi^-)\times {\cal B}(D_0^{*0}\to D^{+}\pi^-)=(6.1\pm0.6\pm0.9\pm1.6)\times10^{-4},
$$
and the  relative phase of the scalar and tensor amplitudes is:
$$
\phi_{D_0^{*0}}=-2.37\pm 0.11\pm0.08\pm0.10.
$$
This is the  first observation of the $D_0^*$.

The $D^*\pi\pi$ final state is described  by the 
production of $D^*_2\pi$, $D'_1\pi$ and $D_1\pi$ with
 $D^{**}\to D^*\pi$.
From a coherent amplitude analysis we obtain the mass, width and product
of the branching fractions for the $D_1$: 
$$
M_{D^{0}_1}=(2421.4\pm1.5\pm0.4\pm0.8)\, {\rm MeV}/c^2,~\Gamma_{D^{0}_1}=(23.7\pm2.7\pm0.2\pm4.0)\, {\rm MeV},
$$
$$
{\cal B}(B^-\to D_1\pi^-)\times {\cal B}(D_1\to D^{*+}\pi^-)=(6.8\pm0.7\pm1.3\pm0.3)\times10^{-4},
$$
and measure the product of the branching fractions for the 
tensor meson process:
$$
{\cal B}(B^-\to D^{*0}_2\pi^-)\times {\cal B}(D_2^{*0}\to D^{*+}\pi^-)=(1.8\pm0.3\pm0.3\pm0.2)\times10^{-4},
$$
and the relative phase of the tensor meson to the axial vector $D^{0}_1$:
$$
\phi_{D^{*0}_2}=-0.57\pm0.14\pm0.06\pm0.13.
$$

We observe the broad $D'_1$ resonance with 
mass and width: 
$$
M_{D'^{0}_1}=(2427\pm26\pm20\pm15)\,{\rm MeV}/c^2,~\Gamma_{D'^{0}_1}=(384^{+107}_{-75}\pm24\pm70)\, {\rm MeV}.
$$
The product of the branching fractions is:  
$$
{\cal B}(B^-\to D'^{0}_1\pi^-)\times {\cal B}(D'^{0}_1\to D^{*+}\pi^-)=(5.0\pm0.4\pm1.0\pm0.4)\times10^{-4}
$$
and the relative phase of $D'^{0}_1$  to $D^{0}_1$ is:
$$
\phi_{D'_1}=1.68\pm 0.20\pm0.07\pm0.16 .
$$

Our analysis also indicates that the axial vector states mix.
The mixing angle is 
$$
\omega=-0.10\pm 0.03\pm0.02\pm0.02,
$$
and the phase is
$$
\psi=0.05\pm0.20\pm0.04\pm0.06.
$$



\section{Acknowledgments}
We wish to thank the KEKB accelerator group for the excellent
operation of the KEKB accelerator.
We acknowledge support from the Ministry of Education,
Culture, Sports, Science, and Technology of Japan
and the Japan Society for the Promotion of Science;
the Australian Research Council
and the Australian Department of Industry, Science and Resources;
the National Science Foundation of China under contract No.~10175071;
the Department of Science and Technology of India;
the BK21 program of the Ministry of Education of Korea
and the CHEP SRC program of the Korea Science and Engineering Foundation;
the Polish State Committee for Scientific Research
under contract No.~2P03B 17017;
the Ministry of Science and Technology of the Russian Federation;
the Ministry of Education, Science and Sport of the Republic of Slovenia;
the National Science Council and the Ministry of Education of Taiwan;
and the U.S.\ Department of Energy.


\begin{thebibliography}{120}
\bibitem{foot1}{The $Z$ coordinate of the track is
defined as the $Z$ coordinate of the track point closest to the beam in
the $r-\phi$ 
plane. $Z$ is the axis opposite to the positron beam direction.}
\bibitem{foot2}{The significance is defined as $\sqrt{-2\ln{{\cal L}_0/{\cal L}_{max}}}$, 
where ${\cal L}_{max}$ is the likelihood with the yield obtained by a fit, and ${\cal L}_0$ is the likelihood 
with the yield constrained to be zero.}.
\end{thebibliography}
\end{document}